\documentclass[twocolumn, twocolappendix]{aastex631}

\usepackage{comment}

\definecolor{indigo}{rgb}{0.0, 0.25, 0.42}

\definecolor{forestgreen}{rgb}{0.13, 0.55, 0.13}

\definecolor{OliveGreen}{rgb}{0,0.6,0}

\newcommand{\Oph}{Oph\,163131}

\shorttitle{A highly settled disk around \Oph}
\shortauthors{Villenave et al.}
\graphicspath{{./}{figures/}}

\begin{document}

\title{A highly settled disk around \Oph}


\email{marion.f.villenave@jpl.nasa.gov}

\author[0000-0002-8962-448X]{M.~Villenave}
\affiliation{Jet Propulsion Laboratory, California Institute of Technology, 4800 Oak Grove Drive, Pasadena, CA 91109, USA}
          \author[0000-0002-2805-7338]{K.~R.~Stapelfeldt }
           \affiliation{Jet Propulsion Laboratory, California Institute of Technology, 4800 Oak Grove Drive, Pasadena, CA 91109, USA}
          \author[0000-0002-5092-6464]{G.~Duch\^{e}ne}
          \affiliation{Astronomy Department, University of California, Berkeley, CA 94720, USA}
          \affiliation{Univ. Grenoble Alpes, CNRS, IPAG, 38000 Grenoble, France}
          \author[0000-0002-1637-7393]{F.~M\'enard}
           \affiliation{Univ. Grenoble Alpes, CNRS, IPAG, 38000 Grenoble, France}
          \author[0000-0001-9321-5198]{M. Lambrechts}
          \affiliation{Lund Observatory, Department of Astronomy and Theoretical Physics, Lund University, Box 43, 22100 Lund, Sweden}
          \affiliation{Centre for Star and Planet Formation, Globe Institute, University of Copenhagen, {\O}ster Voldgade 5–7, 1350 Copenhagen, Denmark}
          \author[0000-0002-5991-8073]{A. Sierra}
          \affiliation{Departamento de Astronom\'ia, Universidad de Chile, Camino El Observatorio 1515, Las Condes, Santiago, Chile}
          \author[0000-0002-8591-472X]{C. Flores }
          \affiliation{Institute of Astronomy, University of Hawaii, 640N Aohoku Place, Hilo, HI 96720, USA}
          \author{W.~R.~F.~Dent}
          \affiliation{Joint ALMA Observatory,Alonso de C\'ordova 3107, Vitacura 763-0355, Santiago, Chile}
          \author[0000-0002-9977-8255]{S. Wolff}
          \affiliation{Steward Observatory, University of Arizona, Tucson, AZ 85721, USA}
          \author[0000-0003-3133-3580]{\'A. Ribas}
          \affiliation{European Southern Observatory, Alonso de C\'ordova 3107, Vitacura, Casilla 19001, Santiago 19, Chile}
          \author[0000-0002-7695-7605]{M.~Benisty}
          \affiliation{Unidad Mixta Internacional Franco-Chilena de Astronom\'ia, CNRS, UMI 3386 Departamento de Astronom\'ia, Universidad de Chile, Camino El Observatorio 1515, Las Condes, Santiago, Chile}
          \affiliation{Univ. Grenoble Alpes, CNRS, IPAG, 38000 Grenoble, France}
          \author[0000-0003-3713-8073]{N. Cuello}
          \affiliation{Univ. Grenoble Alpes, CNRS, IPAG, 38000 Grenoble, France}
          \author[0000-0001-5907-5179]{C. Pinte}
          \affiliation{School of Physics and Astronomy, Monash University, Clayton Vic 3800, Australia}
          \affiliation{Univ. Grenoble Alpes, CNRS, IPAG, 38000 Grenoble, France}

\begin{abstract}

High dust density in the midplane of protoplanetary disks is favorable for efficient grain growth and can allow fast formation of planetesimals and planets, before disks dissipate. Vertical settling and dust trapping in pressure maxima are two mechanisms allowing dust to concentrate in geometrically thin and high density regions. In this work, we aim to study these mechanisms in the highly inclined protoplanetary disk SSTC2D J163131.2-242627 (\Oph, $i\sim$  84$^\circ$). We present new high angular resolution continuum and $^{12}$CO ALMA observations of \Oph. The gas emission appears significantly more extended in the vertical and radial direction compared to the dust emission, consistent with vertical settling and possibly radial drift. In addition, the new continuum observations reveal two clear rings. The outer ring, located at $\sim$100 au, is well resolved in the observations, which allows us to put stringent constraints on the vertical extent of millimeter dust particles. We model the disk using radiative transfer and find that the scale height of millimeter sized grains is 0.5\,au or less at 100\,au from the central star. This value is about one order of magnitude smaller than the scale height of smaller micron-sized dust grains constrained by previous modeling, which implies that efficient settling of the large grains is occurring in the disk. When adopting a parametric dust settling prescription, we find that the observations are consistent with a turbulent viscosity coefficient of about~$\alpha\lesssim10^{-5}$ at 100\,au. 
Finally, we find that the thin dust scale height measured in \Oph\ is favorable for planetary  growth  by pebble accretion: a 10\,M$_{\rm E}$ planet may grow within less than 10\,Myr, even in orbits exceeding 50~au. %


\end{abstract}


\section{Introduction} 
\label{sec:intro}

In recent years, a number of studies presented direct or indirect detections of planets embedded in protoplanetary disks~\citep[e.g.,][]{Keppler_2018, Pinte_2018, Teague_2018}, indicating that planet formation is a fast process. In the core accretion paradigm, micron-sized particles that are inherited from the interstellar medium need to grow quickly to form a planetary core which can attract the surrounding gas before the disk has dissipated. 
The streaming instability~\citep[][]{Youdin_2005} is one of the currently favored mechanisms allowing a sufficiently rapid growth, from pebbles to planetesimals, in the disk phase. For this instability to develop, the dust-to-gas ratio needs to be high in the disk midplane, of the order of 1, and large pebble-sized particles need to be present. After this planetesimal formation stage, planetary embryos can continue to form the cores of giant planets by accreting remaining inwards-drifting pebbles \citep{Lambrechts_2012}.

Various studies have shown that substructures, in particular rings and gaps, are  ubiquitous in protoplanetary disks in the Class II phase~\citep{Long_2018, Huang_2018}, and already present in some young Class~I disks~\citep{ALMA_HLTau_2015, Segura-Cox_2020}. In many cases, they act as dust traps and grain growth is associated with rings~\citep[e.g.,][]{Carrasco-Gonzalez_2019,Macias_2021, Sierra_2021}. On the other hand, the efficiency of dust vertical settling remains largely unconstrained. Similarly to radial drift~\citep{Weidenschilling_1977}, this mechanism is a balance between the stellar gravity and the interaction of dust with gas. Large grains (e.g., millimeter sizes) are expected to settle most efficiently towards the midplane, while smaller grains (e.g., micron sizes) are predicted to remain well coupled to the gas and are co-located with it. At the same time, because of radial drift, large grains are also predicted to drift towards the star. 

Currently the vertical scale height of the gas in protoplanetary disks around T Tauri stars has been constrained in a number of studies to be about 10\,au at a radius of 100\,au~\citep[e.g.,][]{Burrows_1996, Watson_2007, Wolff_2017}.  
This was mainly done by modeling scattered light images, which probe small dust grains assumed to be well coupled to the gas. Some other studies also estimated near-infrared scattering surface extent~\citep[e.g.,][]{Ginski_2016, Avenhaus_2018} or the height of gas emission layers~\citep[e.g.][]{Pinte_2018, Law_2021, Rich_2021}, using less model-dependent techniques. However, these latter techniques do not probe the pressure  scale height of the disk but the scattering or emission surfaces, which may be several times higher than the physical scale height. 

On the other hand, only a few studies have estimated the scale height of millimeter dust particles, most affected by vertical settling. This is in part because the majority of observed protoplanetary disks are moderate inclination systems in which it is difficult to constrain the thin vertical extent of millimeter grains. 
For these systems detailed modeling of gaps and rings is needed~\citep[e.g.,][]{Pinte_2016, Doi_Kataoka_2021}. On the other hand, disks seen edge-on offer the most favorable orientation to study their vertical extent. \citet{Villenave_2020} presented the first high angular resolution ($\sim$0.1\arcsec) millimeter survey of edge-on disks. By comparing the vertical and radial extent of  3  systems with radiative transfer models, they showed that the scale height of the millimeter dust particles is of a few au at 100 au, significantly smaller than the typical gas scale height. These results indicate that efficient vertical settling is occurring in protoplanetary disks. Further studies are needed to increase the number of disks with known gas and millimeter dust vertical extent in order to better understand the efficiency of vertical settling.

In this paper, we focus on SSTC2D J163131.2-242627 (hereafter \Oph), a highly inclined protoplanetary disk located in the Ophiuchus star-forming region, at a distance of about $147\pm3$\,pc~\citep{Ortiz-Leon_2017, Ortiz-Leon_2018}. \Oph\ was included in the survey of \citet{Villenave_2020}, and has recently been studied by two companion papers~\citep{Flores_2021, Wolff_2021}, which presented ALMA observations at an angular resolution of $\sim 0.2\arcsec$, as well as scattered light HST and Keck images. \citet{Wolff_2021} used radiative transfer to model both the 1.3\,mm ALMA image and the 0.8\,$\mu$m HST images with an extensive MCMC framework. They found that some degree of vertical settling is needed to reproduce both observations. On the other hand, \citet{Flores_2021} analyzed the ALMA $^{12}$CO and $^{13}$CO maps to characterize the 2D temperature structure of the disk. They obtained a dynamical mass estimate of $1.2\pm0.2~M_\odot$ from the ALMA observations, and characterized the spectral type of the source to be K4, using optical and near-infrared spectroscopy.  

In this work, we present new high angular resolution observations of \Oph\  at 1.3\,mm (resolution of $\sim 0.02\arcsec$ or 3\,au). The new images reveal a number of rings that were not detected with previous millimeter observations. We derive a detailed radiative transfer model of the new millimeter observations to characterize the physical structure of the disk and add constraints on vertical settling. In Sect.~\ref{sec:observations}, we present the observations and data reduction. In Sect.~\ref{sec:results}, we present the main features seen in the images. Then, in Sect.~\ref{sec:model} we describe the disk model and present the results. We discuss the implication of our results in Sect.~\ref{sec:discussion}. Finally the conclusions are presented in Sect.~\ref{sec:concl}.

\begin{figure*}
\centering
    \includegraphics[width = 0.49\textwidth]{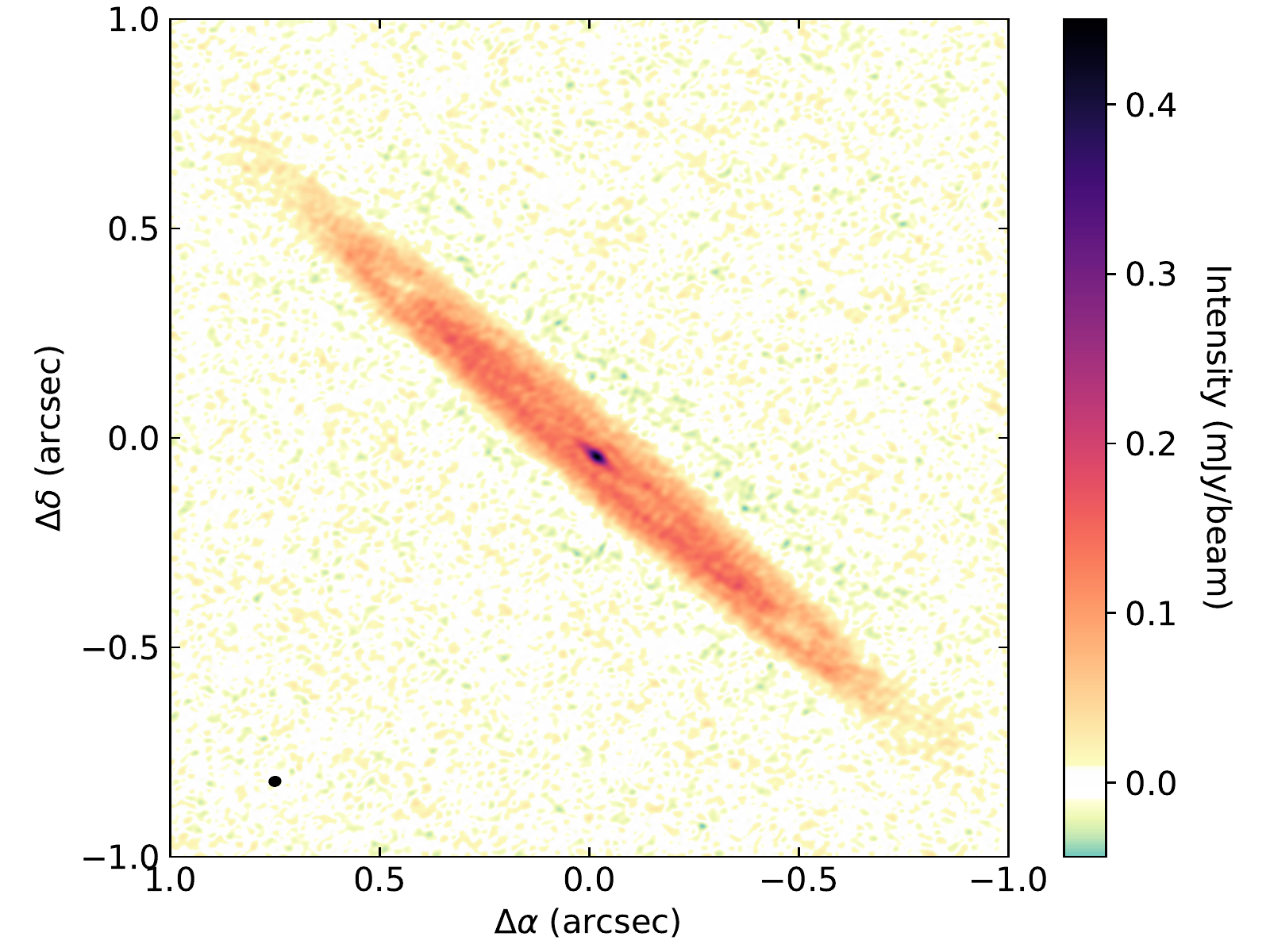}
    \includegraphics[width = 0.49\textwidth]{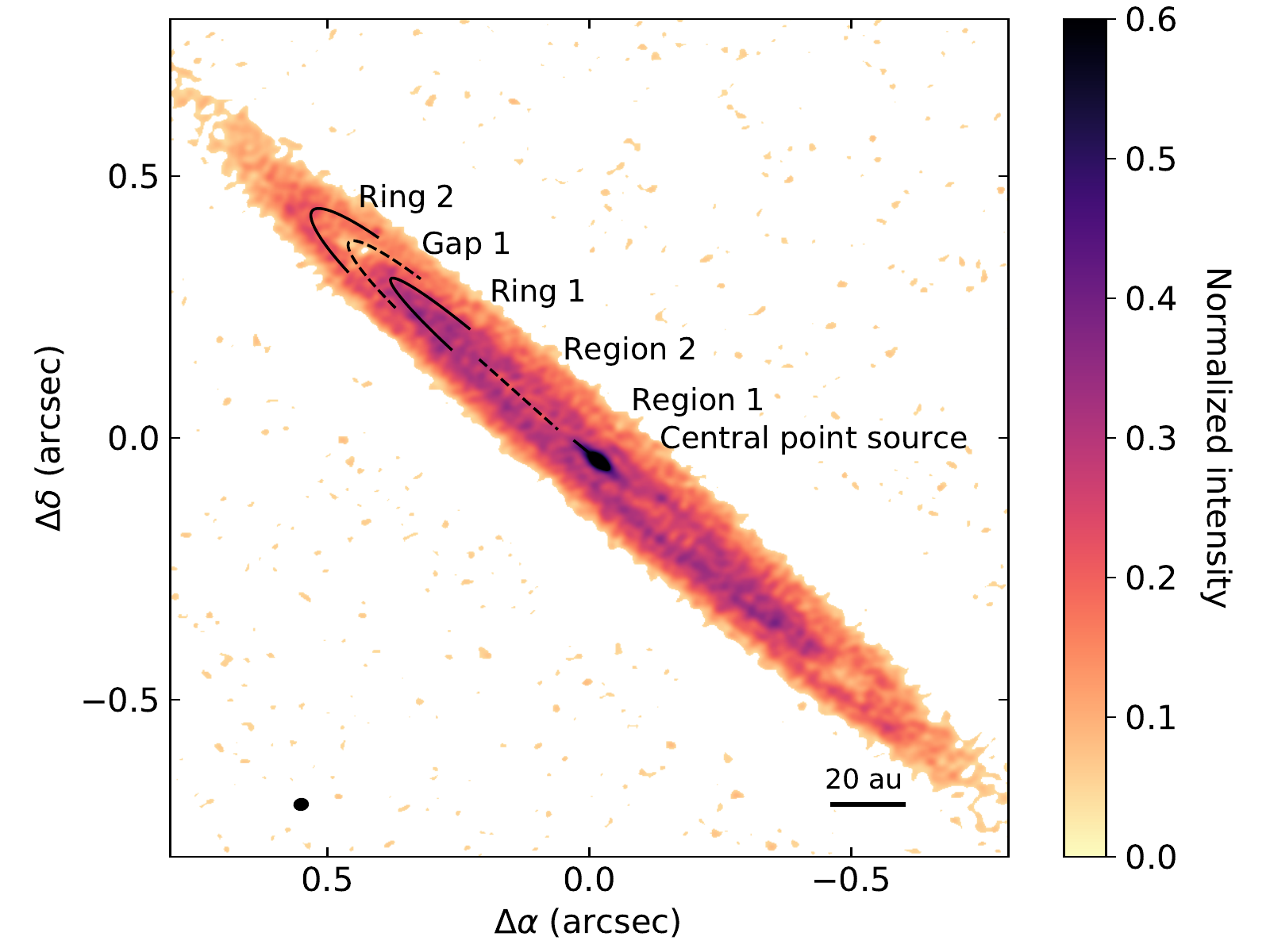}
    \caption{\emph{Left:} Continuum image of \Oph. The beam size is indicated by an ellipse in the bottom left corner of the plot. North is up and east is left.  \emph{Right:} Labeled zoomed-in continuum image of \Oph, scaled so that the rings are more visible~(see Sect.~\ref{sec:model_ALMA} and Fig.~\ref{fig:model_1330_cut_zoom} for details on the identification of region~1). The central region is saturated (black) and pixels with less than 2$\sigma$ appear in white.  The beam size is indicated by an ellipse in the bottom left corner of the plot.  }
    \label{fig:cont}
    \label{fig:cont_label}
\end{figure*}

\section{Observation and data reduction}
\label{sec:observations}

We present new cycle~6 observations of \Oph\  in band~6~(1.3\,mm, Project: 2018.1.00958.S, PI: Villenave). The spectral setup was divided in three continuum spectral windows of rest frequency 229.0\,GHz, 243.5\,GHz, and 246.0\,GHz, and a fourth spectral window including the $^{12}$CO $J=2-1$ transition at 230.538\,GHz. 
The line spectral window has a native velocity resolution of 0.64\,km\,s$^{-1}$. 
The data were obtained on June 8, 2019, with baselines ranging from 80\,m to 16\,km. The total observing time on source was about 2~hours and 30~minutes. The raw data were calibrated using the CASA pipeline version~5.4.0~\citep{McMullin_2007}, and the rest of the data processing was done with CASA~5.6.1.


To produce the final images, we combined our cycle~6 observations with lower angular resolution observations from cycle~4~(Project 2016.1.00771.S, PI: Duch\^ene) previously published \citep{Villenave_2020, Flores_2021, Wolff_2021}. We note that a shift of about 0.05\arcsec~($\sim$20\% of the cycle~4 beam) was present between the cycle~4 and cycle~6 observations, which is roughly consistent with the astrometric accuracy of ALMA~(see ALMA technical handbook\footnote{https://almascience.eso.org/documents-and-tools/cycle8/alma-technical-handbook}). Thus, before producing the combined image, we aligned both observations using the \texttt{fixplanet} CASA task (with the option \texttt{fixuvw} as True).  To maximize the dynamic range of the final image, we performed phase self-calibration on the cycle~4 continuum observations as mentioned in \citet{Flores_2021, Villenave_2020}. Due to the limited signal-to-noise per beam, no self-calibration could be performed on the cycle~6 observations, nor on the combined cycle~4 and cycle~6 data. We then produced the continuum image using the CASA \texttt{tclean} task on the combined dataset, with a Briggs weighting (robustness parameter of +0.5) and using the multiscale deconvolver, with scales of 0, 1, 5, and 10 times the beam FWHM. The resulting continuum beam size is 0.024\arcsec$\times$0.020\arcsec ($\sim3.5\times3$\,au) with a major axis at PA\,$=-81^\circ$, and the resulting continuum rms is 9.3~$\mu$Jy/beam.

After applying the continuum self-calibration solutions to all spectral windows, we subtracted the continuum emission using the \texttt{uvcontsub} task in CASA. Then, we derived the emission line maps from the calibrated visibilities using the \texttt{tclean} function. We use 0.7\,km\,s$^{-1}$ velocity resolution and a Briggs weighting~(robustness parameter of +0.5) to create the line images. Additionally, to increase the signal to noise of the $^{12}$CO observations, we applied a uv-taper while generating the images. The combined resulting $^{12}$CO beam is  0.081\arcsec$\times$0.072\arcsec\ with a major axis at PA\,$=89^\circ$, and the average rms of the line is 0.8~mJy/beam per channel. Finally, we generate moment~0 and 1 maps including only pixels above 3~times the rms. 

\section{Results}
\label{sec:results}

\subsection{Continuum emission}
\label{sec:frank}

We present the continuum observations of \Oph\, in~Fig.~\ref{fig:cont}. The high angular resolution of the image reveals several rings even though the disk is highly inclined. We highlight the structures in the right panel of Fig.~\ref{fig:cont}. 
From outside in, we see two rings (ring~2 and ring~1) separated by a clear gap (gap~1), some emission inside of the second ring (that we call region~2 in Fig.~\ref{fig:cont}), and an inner central emission (region~1 and central point source, see  Sect.~\ref{sec:model_ALMA}). We note that the existence of the outer gap was hinted by a shoulder detected in previous observations~\citep{Villenave_2020}, but it was not resolved as with the current image.

To characterize the substructures, we fit the 
deprojected visibilities using the \texttt{frank} package~\citep{Jennings_2020}.
We exported the visibilities using a modified version of \texttt{export\_uvtable} from \citet{uvplot_mtazzari}. In our version, instead of using the average wavelength from all data, each baseline is normalized using the wavelength associated to each spectral window and channel. The visibilities from cycle~6 and~4 are independently obtained, shifted, and deprojected using an inclination and position angle of 84$^\circ$ and $49^\circ$, respectively (see Sect.~\ref{sec:model}). Then, they are concatenated.
The radial profile obtained from \texttt{frank} depends on two hyper-parameters: $\alpha_{frank}$, $w_{\rm smooth}$. The former controls the maximum baseline that is used in the fitting process (long baselines where the signal-to-noise ratio is small are not included), and the latter controls how much the visibility model is allowed to fit all the visibilities structures from the data (see \citealt{Jennings_2020} for more details).  We ran 1000 fits using different combinations of these two hyper-parameters varying between $1.2 < \alpha_{frank} < 1.4$ and $-2 <\log_{10} (w_{\rm smooth})< -1$. We chose these ranges to avoid artificial high frequency structure, and observe no significant difference between the profiles. In addition, we chose not to allow the resulting radial profiles from \texttt{frank} to have negative values. 

\begin{figure}
\centering
    \includegraphics[width = 0.48\textwidth, trim={1cm 0cm 0cm 0cm}, clip]{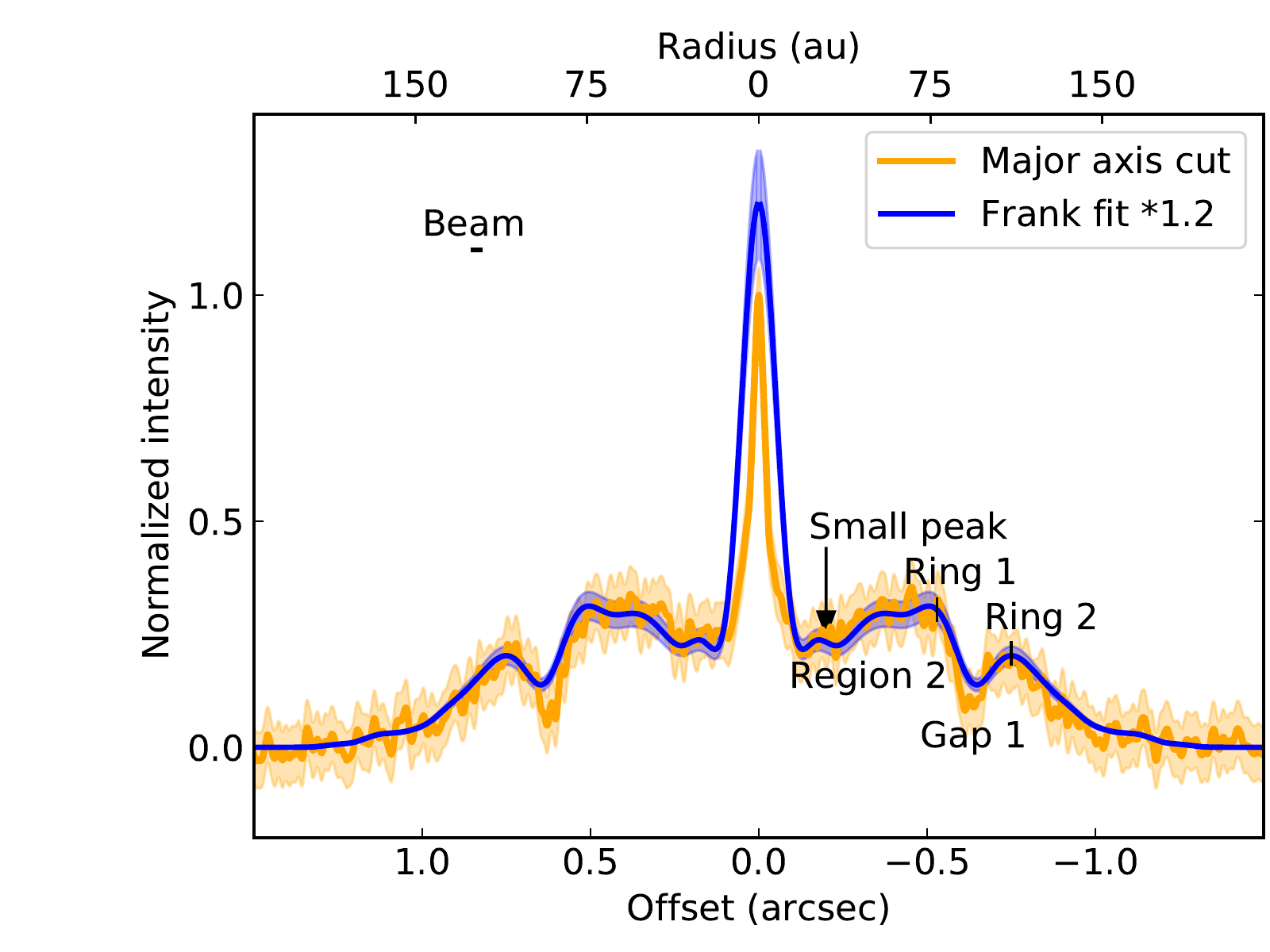}
    \includegraphics[width = 0.48\textwidth, trim={1cm 0cm 0cm 1cm}, clip]{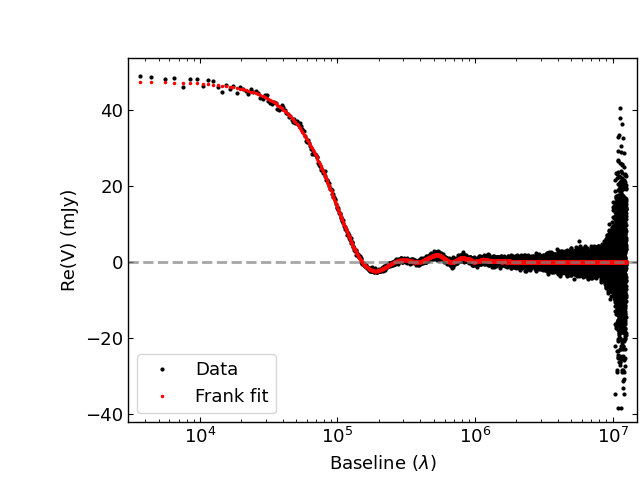}
    \caption{\emph{Top:} Mean radial intensity profile of millimeter continuum emission obtained from the visibilities with \texttt{frank} (in blue), and major axis cut obtained from the image (in orange). The major axis cut is normalized to its peak and the \texttt{frank} profile is normalized so that the normalized intensity of ring~2 coincides in both profiles. The uncertainties on the visibility fit correspond to the typical 10\% flux calibration error which dominates over the fits uncertainty, while the uncertainty on the major axis cut corresponds to $\pm3\sigma$. \emph{Bottom:} Observed averaged deprojected visibilities (black) and model obtained with \texttt{frank} (red). }
    \label{fig:frank}
\end{figure}

The average resulting radial profile and the average real part of the 1000 visibility fits are presented in Fig.~\ref{fig:frank}. In the top panel, we also include the cut along the major axis obtained from the image~(Fig.~\ref{fig:cont}).
We find that the structures obtained from the visibility fit and the major axis cut are in very good agreement. They share the same location, and the relative brightness between ring~1 and ring~2 is similar in both cases. Gap~1 appears deeper in the major axis cut than in the visibility fit, likely because the visibility fit is averaged over all angles and the gaps are filled along the minor axis due to projection effects.

Starting from the outside, we find that ring~2 peaks at $0.73\pm0.02$\arcsec, gap~1 is lowest at $0.63\pm0.02$\arcsec, and ring~1 peaks at $0.51\pm 0.02$\arcsec. Inward of $\sim$0.5\arcsec, there is a flat region with increasing surface brightness with radius, that we name region~2. In the \texttt{frank} profile, we also detect a small peak at $\sim$0.18\arcsec\ (inside of region~2), that is not visible in the major axis cut. Finally, we detect some emission from the inner regions of the disk. In the major axis cut, the central emission can be described by a central marginally resolved Gaussian plus some extended emission (possibly a ring, see Sect.~\ref{sec:model_ALMA} and Fig.~\ref{fig:model_1330_cut_zoom}). The two components are not detectable in the visibility fit likely because this is an azimuthally averaged profile and the inner ring is not resolved in the minor axis direction, even in the visibility domain. In addition, we find that the central flux of the \texttt{frank} fit is greater than that of the major axis cut. This is because, contrary to the major axis cut, the visibility fit is an azimuthal average, which is differently affected by the central beam smearing. 

\begin{figure*}
    \centering
    \includegraphics[width = 0.48\textwidth, trim={1cm 0cm 0.5cm 0cm}, clip]{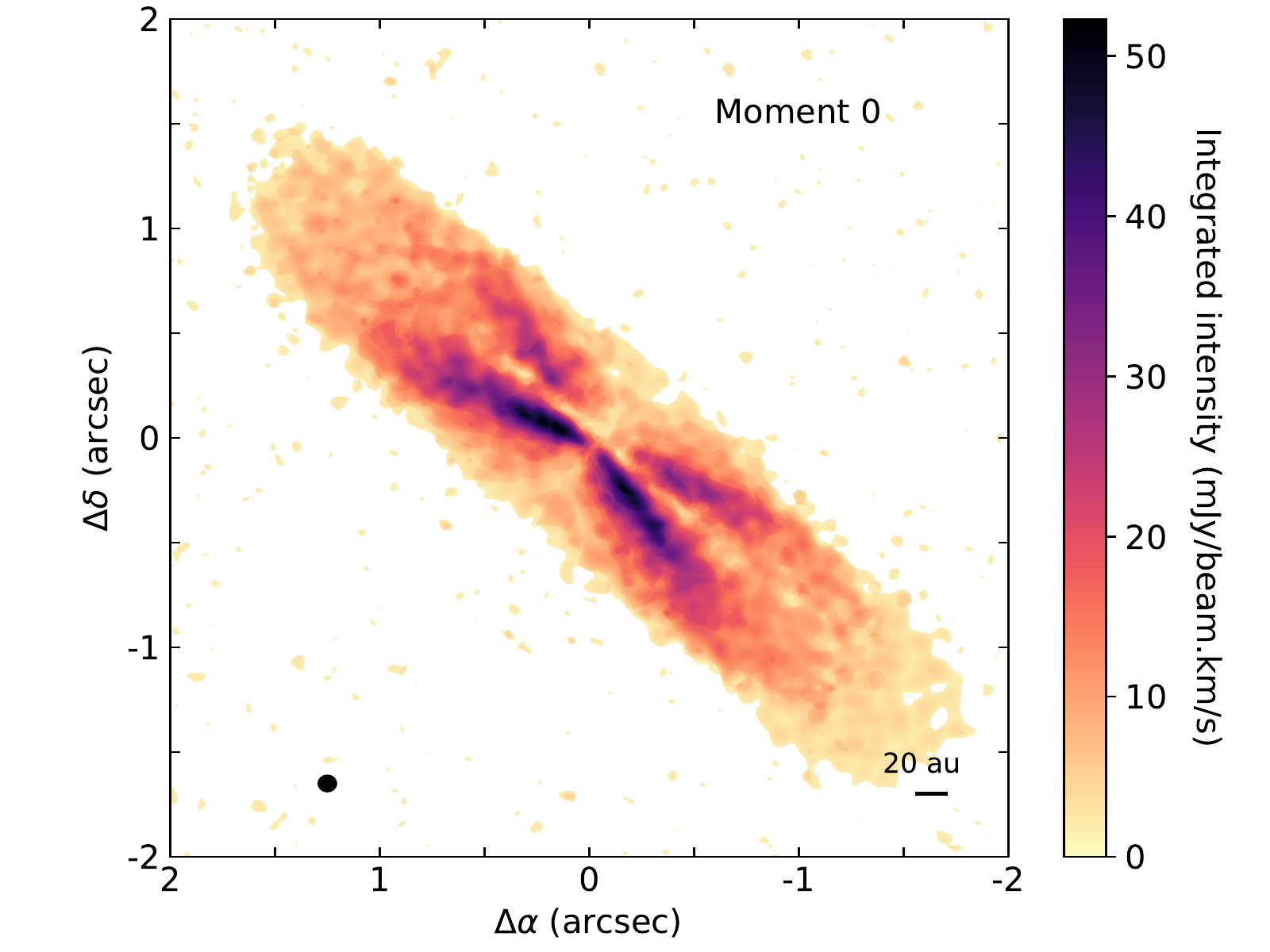}
    \hspace{0.5cm}
    \includegraphics[width = 0.48\textwidth, trim={1cm 0cm 0.5cm 0cm}, clip ]{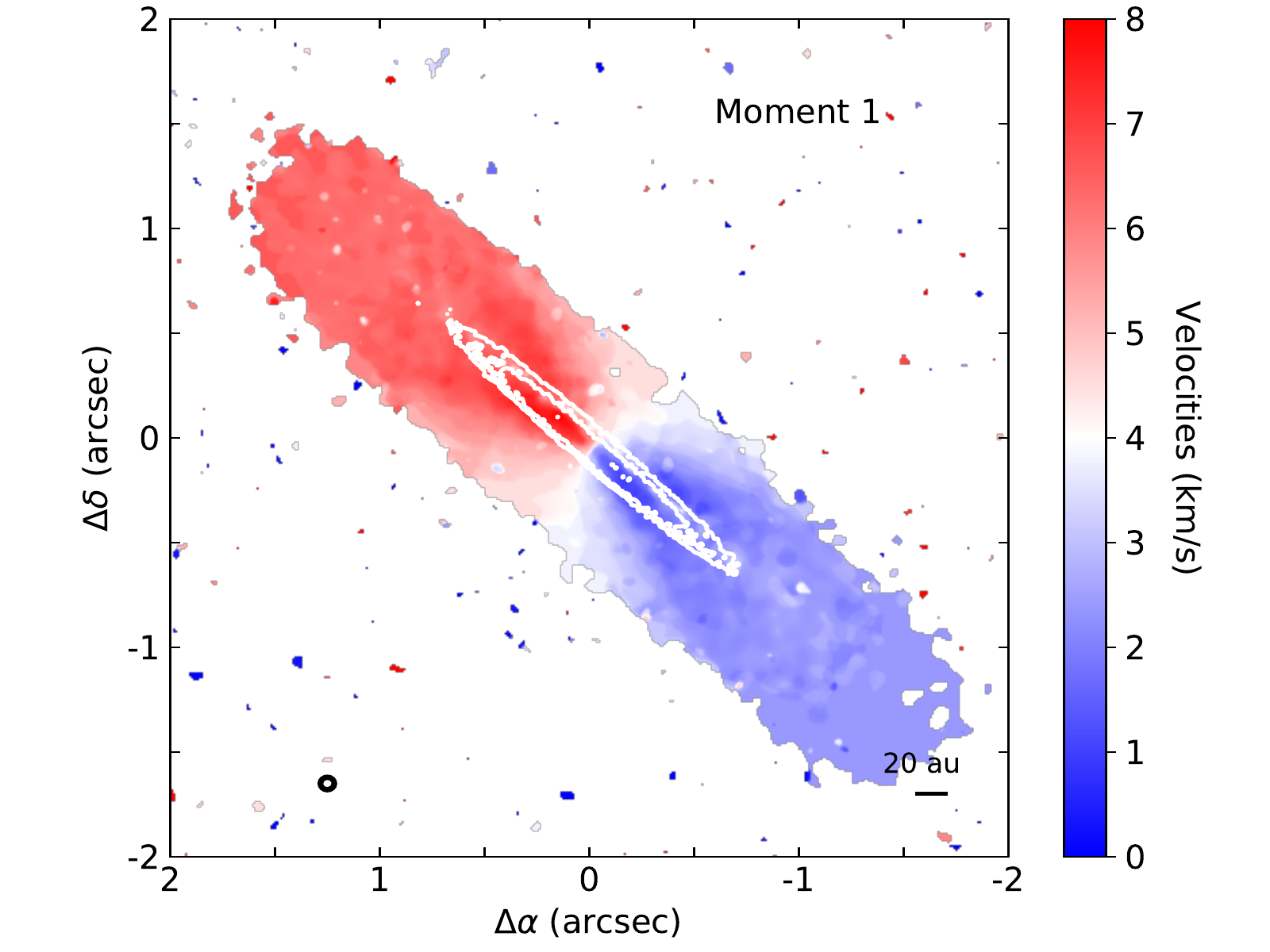}
    \caption{\emph{Left:} $^{12}$CO continuum substracted moment 0 map, \emph{Right:} $^{12}$CO  moment 1 map and continuum $5 \sigma$ and $10 \sigma$  contours in white. We illustrate the beam sizes by ellipses in the bottom left corner of each panel ($^{12}$CO beam in black -- both panels --, and continuum beam in white within the CO beam -- right panel only). }
    \label{fig:CO}
\end{figure*}

In Sect.~\ref{sec:model}, we present a radiative transfer modeling of the continuum millimeter emission. By reproducing the features characterized with the visibility fit, we aim to constrain the physical structure of the disk and of the different grain populations. In particular, we use the presence and shape of gap~1 to constrain the vertical extent of millimeter sized particles in the disk. As both the \texttt{frank} profile and major axis cut show similar substructures, we use the major axis profile for the rest of the analysis. 

\subsection{CO emission}
\label{sec:CO_data}

We display the $^{12}$CO moment 0 and 1 maps in Fig.~\ref{fig:CO}. Those  represent the integrated intensity and the velocity map, respectively. Before discussing the shape of the maps, we first estimate the integrated flux of the $^{12}$CO emission, using the CASA \texttt{imstat} function on our ALMA moment~0 map. We measure the flux over a rectangle of 0.9\arcsec\ along the minor axis direction and 4.4\arcsec\ along the major axis, and obtain $F_{^{12}CO} = 6.1 \pm 0.6$\,Jy\,km\,s$^{-1}$. We report a 10\% uncertainty due to flux calibration errors.

From  the overlay of the $^{12}$CO moment 1 map and continuum emission in Fig.~\ref{fig:CO}, we see that the  $^{12}$CO emission is significantly more extended than the millimeter continuum emission, both in the radial and vertical directions~\citep[see also][]{Flores_2021}. This is consistent with large grains being affected by vertical settling~\citep{Barriere-Fouchet_2005} and possibly radial drift~\citep{Weidenschilling_1977}. 

Thanks to the increased angular resolution, the $^{12}$CO moment 0 map (see left panel of Fig.~\ref{fig:CO}) shows significantly more details than previously detected in \citet{Flores_2021}, though the same overall features are present. The bottom south-east side appears brighter than the north-west side of the disk, which indicates that the south-east side is closest to us~\citep[see also][]{Wolff_2021}. In addition, the disk emission can be described by two distinct regions. First, the inner 150\,au ($\sim 1\arcsec$) of the disk describe a X shape, typical of very inclined systems. With the new observations, we clearly resolve the cold midplane with little CO emission, which separates the two bright sides of the disk. This inner region is colocated with the continuum millimeter emission. On the other hand, at larger radii ($R>200$\,au, $1.4\arcsec$), the disk appears flat, with a nearly constant brightness profile with radius.  
We find that the transition from the inner flaring X-region to the outer flat region starts at a radius of $\sim$150\,au, which roughly corresponds to the radius where the millimeter  continuum emission disk ends and the disk' scattered light stops (see Sect.~\ref{sec:HST} and Fig.~\ref{fig:overlay_hst}). This change in the dust structure seems to also corresponds to a change in the disk gas structure.

Using lower angular resolution observations, \citet{Flores_2021} found that this outer region is associated with a uniform temperature both radially and vertically. In Appendix~\ref{appdx:TRD}, we present an updated temperature map of the disk, resulting from the Tomographically Reconstructed Distribution method \citep[TRD,][]{Flores_2021, Dutrey_2017} applied to the new high angular resolution observations. We also recover the isothermal outer region. We note that \citet{Flores_2021} suggested that it was due to external UV illumination through a mostly optically thin region, and refer the reader to their analysis for a more detailed interpretation. 

\subsection{Comparison with scattered light image}
\label{sec:HST}

\begin{figure}
    \centering
    \includegraphics[width = 0.49\textwidth]{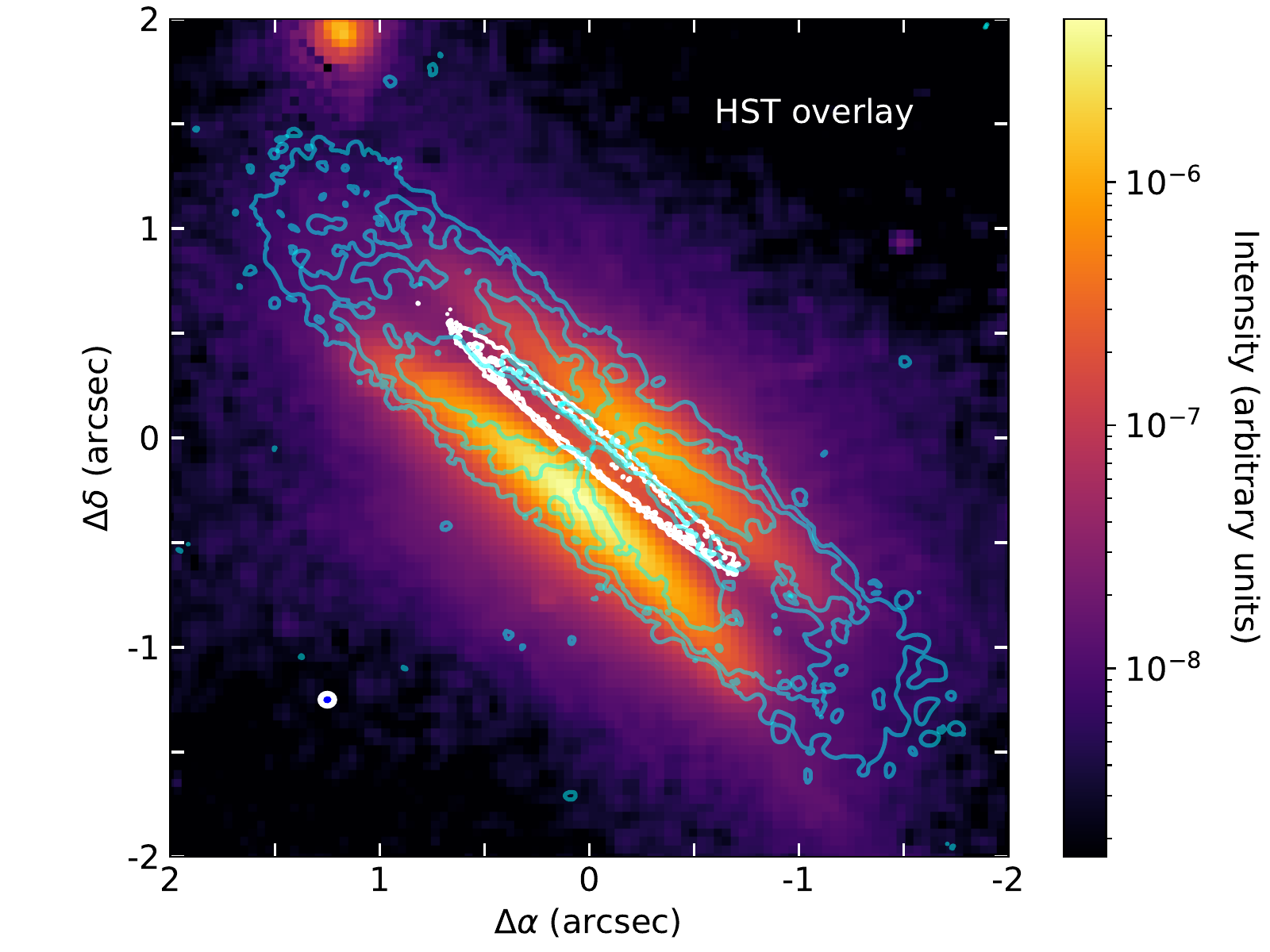}
    \caption{HST 0.6 $\mu$m scattered light image (colors), $^{12}$CO moment 0 ($5 \sigma_{gas}$, $15 \sigma_{gas}$, and $30 \sigma_{gas}$  contours, in blue), and $5 \sigma$ and $10 \sigma$  contours of the continuum map (white).  The beam sizes are shown by ellipses in the bottom left corner ($^{12}$CO beam in white, and continuum beam in black, inside the CO beam).}
    \label{fig:overlay_hst}
\end{figure}

In this section, we compare the high angular resolution millimeter continuum and $^{12}$CO observations with a HST scattered light image of \Oph\ modeled in \citet{Wolff_2021}. The scattered light image is expected to trace the scattering surface of small micron-sized particles, well mixed with the gas.  We present an overlay of the HST image at 0.6\,$\mu$m, the $^{12}$CO moment 0 emission, and millimeter continuum contours in Fig.~\ref{fig:overlay_hst}. As first shown in  \citet{Stapelfeldt_2014}, the scattered light image of \Oph\ presents the characteristic features of edge-on disks, with two parallel nebulosities separated by a dark lane. \citet{Villenave_2020} estimated that the disk size~(diameter) in scattered light is 2.5\arcsec. 

We first compare the apparent radial and vertical extent of the disk in the different tracers. In the radial direction, we find that the millimeter dust emission is less extended than both the scattered light and $^{12}$CO emission. The outer radius of the disk millimeter dust emission (at $\sim$150\ au) appears to coincide with the 15$\sigma_{gas}$ contours which also marks the limit of the inner region of $^{12}$CO emission (characterized by the bright X pattern split by a cold midplane, see Sect.~\ref{sec:CO_data}). Further away, the scattered light is fainter and comes from a more diffuse region which seems to extent as far out radially as the gas emission (out to approximately 400~au). The change of intensity in scattered light suggests that small grains are either not illuminated by the central star, possibly due to a decrease of the height of their scattering surface, or that they are depleted in the outer regions of the disk~\citep[e.g.,][]{Muro-Arena_2018}.  

Both the scattered light and the $^{12}$CO emission appear significantly more extended vertically than the millimeter continuum image, which is consistent with vertical settling occurring in the disk. The disk in scattered light  seems to be as extended vertically as the $^{12}$CO emission, but with an additional fainter halo extending to significantly higher levels, which is due to PSF convolution (see for example the vertical halo also present in the PSF-convolved model of the HST image in Appendix~\ref{appdx:SED_HST}).

\begin{figure}
    \centering
    \includegraphics[width = 0.46\textwidth, trim={0cm 0cm 0cm 0cm}, clip,]{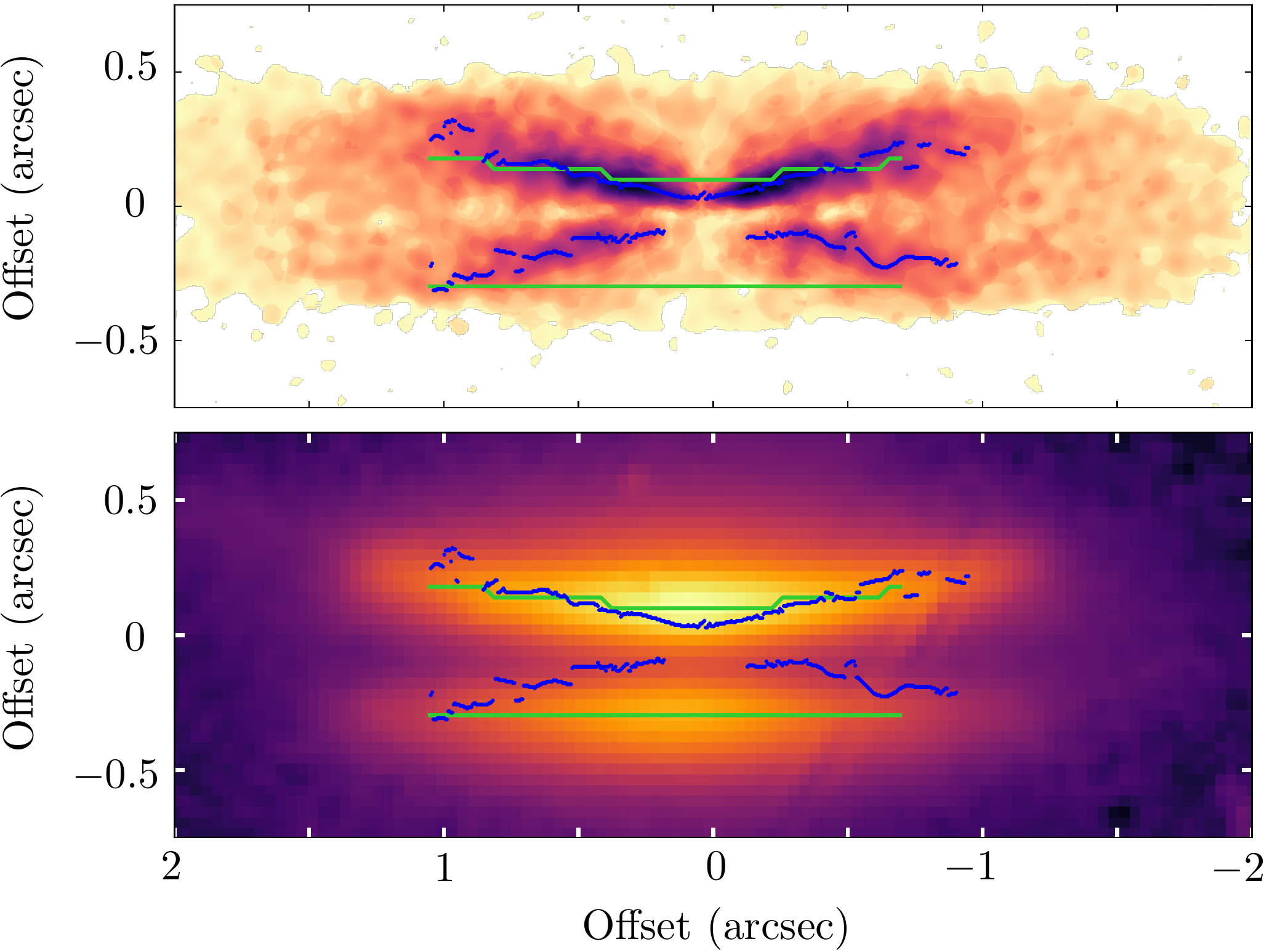}
    \caption{Position of the spines of the 0.6$\mu$m image (green lines), and $^{12}$CO moment 0 maps (blue lines), overlayed on top of either the moment 0 map (top panel) or the scattered light image (bottom panel).}
    \label{fig:spines}
\end{figure}

Interestingly, we also see significant differences between the morphologies seen in scattered light and $^{12}$CO emission. In Fig.~\ref{fig:spines}, we applied the method described in Appendix~D of \citet{Villenave_2020} to show the spine  of each nebula (peak intensity) as seen in scattered light and $^{12}$CO emission. We find that, up to about 150 au, there is a clear increase of the $^{12}$CO apparent height with radius (blue lines). When deprojected, \citet{Flores_2021} found that this increase is roughly linear. We fitted the vertical extent of the $^{12}$CO emission surface as a function of radius with a linear regression and obtain $z/r\sim0.3$, for $r<150$~au. The variation of the CO scattering surface with radius in \Oph\ is relatively small compared to similar estimates in other protoplanetary disks~\citep[e.g.,][]{ Pinte_2018, Flores_2021, Law_2021}.
On the other hand, the opposite is seen for the scattered light nebulae (green lines), which are extremely flat, in the sense that their apparent height does not vary significantly with projected distance. This difference in behavior is particularly clear in the least bright side of the disk, which is in the bottom of Fig.~\ref{fig:spines}, and is likely due to optical depth effects.

Differences between the height of the dust scattering surface and CO emission surfaces have already been identified in other protoplanetary disks~\citep[e.g.,][]{Rich_2021}, and can be due to different optical depth in the different tracers. In the case of our study, we are comparing the integrated scattered light intensity to the $^{12}$CO moment map, which is the integration of the $^{12}$CO channel maps. In the channel maps, higher velocities detected at small projected distances allow us to probe the warmer (brighter) disk interior. On the other hand, scattered light is optically thick and comes from the disk outer edge even at small projected distances, as indicated by its non-variation with projected distance. Thus, at small projected distances it is expected that the scattered light surface, probing the disk outer edge, appears higher than the $^{12}$CO moment~0 peak of emission, which traces the inner regions.  The agreement in height towards the outer radii however indicates that both components are emitted at similar altitudes above the midplane.

\section{Radiative transfer modeling}
\label{sec:model}

In the previous section, we presented high angular resolution observations of \Oph, which reveal a highly structured disk. Given the high inclination of the disk, the presence of rings provides constraints on the vertical extent of millimeter dust particles, and on the efficiency of vertical settling. We aim to use these additional morphological constraints to refine the millimeter continuum radiative transfer modeling of the source presented in \citet{Wolff_2021}, and to constrain the physical structure of the millimeter grains in the disk of \Oph. Building on the existing model and to obtain a more complete view of the disk, we also compute the spectral energy distribution (SED) and the 0.6 $\mu$m image.

\subsection{Methodology}
\label{sec:mcfost}

To model the disk of \Oph, we use the radiative transfer code \textsc{mcfost}~\citep{Pinte_2006, Pinte_2009}. 
We assume that the disk structure is axisymmetric, and we model the disk using several regions to reproduce all substructures seen in the continuum observations. Given the complexity of the ALMA continuum image, we do not aim to find a unique model, but rather a well-fitting one.   

For each region, we assume a power-law distribution for the surface density:
	\begin{equation}
		\label{eq:surf_dens}
		\Sigma(r) \propto r^{p}, \text{ for } R_\mathrm{in} < r < R_\mathrm{out}
	\end{equation}
We parametrize the vertical extent of the grains by the scale height, such that:
	\begin{equation}
		\label{eq:scale_height}
		H(r) = H_\mathrm{100\,au} (r/100\,\text{au})^\beta, \text{ for } R_\mathrm{in} < r < R_\mathrm{out}
	\end{equation}
where $\beta$ is the flaring exponent. For simplicity, we use astronomical silicates \citep[similar to those shown in Fig.~3 of][]{Draine_1984}\footnote{The dust properties are evaluated from: ftp://ftp.astro.princeton.edu/draine/dust/diel/eps\_suvSil}
with a power law distribution of the grain size following $n(a)\text{d}a \propto a^{-3.5}\,\text{d}a$. 

The main free parameters of each region are thus $R_\mathrm{in}$, $R_\mathrm{out}$, $H_\mathrm{100\,au}$, and $M_\mathrm{dust}$. In all our modeling we fixed a flaring exponent of $\beta=1.1$ for large grains, and $\beta=1.2$ for small grains~\citep[as in][for example]{Villenave_2019}. 
In addition, we fixed the surface density exponent to $p=-1$ in all regions, except for region~2 and ring~2 where we needed to adjust this exponent~(see Sect.~\ref{sec:model_ALMA}). 
The inclination is a global disk parameter that we constrain with this modeling to be 84$^\circ$ (see Sect.~\ref{sec:model_ALMA}, ring~2). We assume that all regions are coplanar with this inclination.  
Following \citet{Wolff_2021} and \citet{Flores_2021}, we adopt a distance of 147~pc, a stellar mass of 1.2~M$_\odot$, stellar radius of 1.7~R$_\odot$, stellar effective temperature of 4500~K, and assume 2 magnitudes of interstellar extinction in the SED. 

For each set of parameters, we compute the 1.3\,mm continuum image, SED, and 0.6\,$\mu$m scattered light image. All maps are convolved by a representative PSF before being compared to the data. We use the HST PSF generated by the \texttt{TinyTim} software package~\citep{Krist_2011} for the 0.6\,$\mu$m image. For the ALMA data, we simulate the real interferometric response by producing synthetic images using the CASA simulator. For each model, we first generate synthetic visibility files for all 3 observing times and configurations, using the CASA task \texttt{simobserve}. We then produce synthetic images of these visibilities using the \texttt{tclean} function with the same weighting parameters as for the observations.\\

\begin{figure*}
    \centering
    \includegraphics[width = 0.3\textwidth, trim={0cm 0cm 3.2cm 0cm}, clip ]{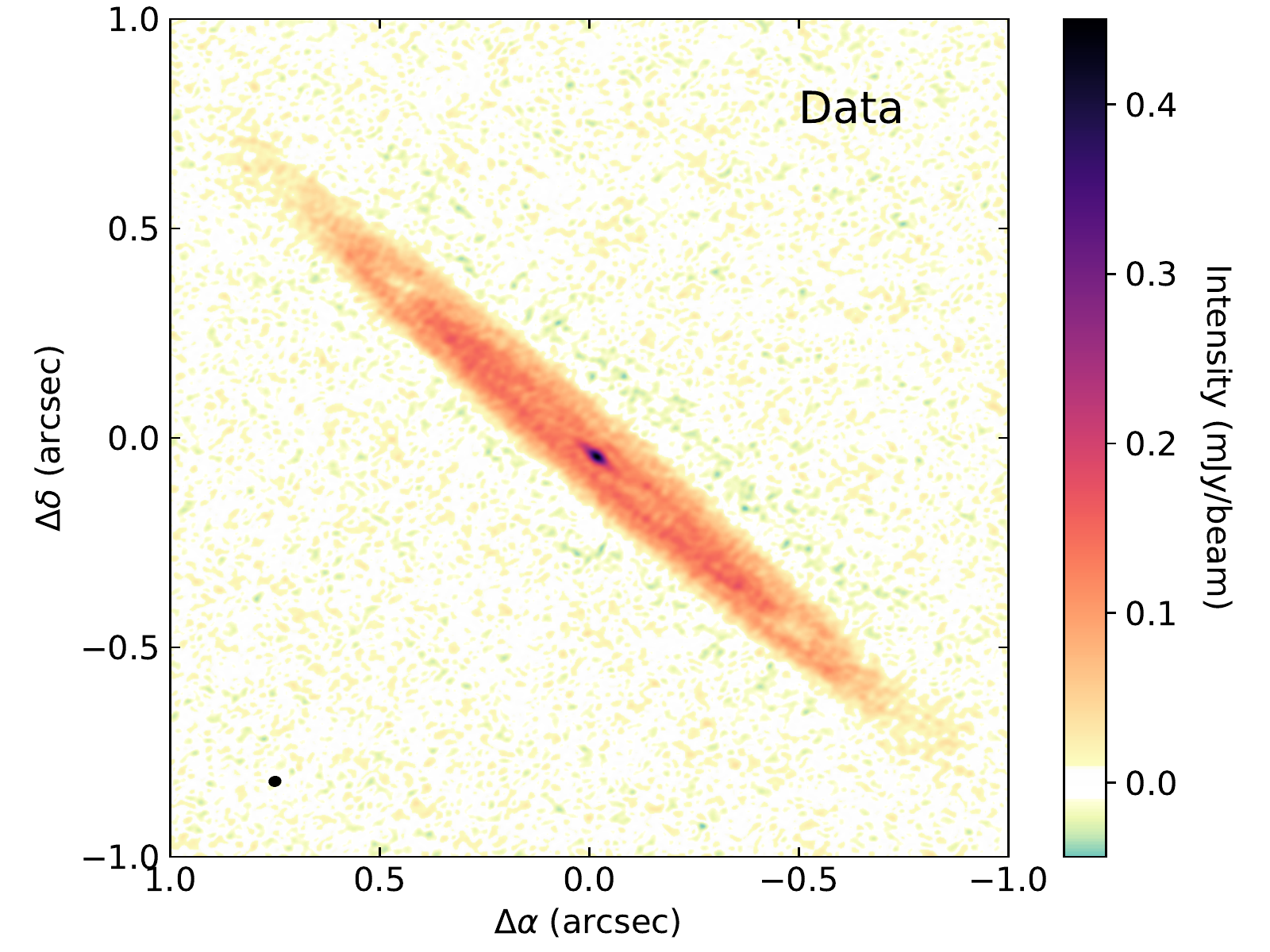}
    \includegraphics[width = 0.3\textwidth, trim={0cm 0cm 3.2cm 0cm}, clip ]{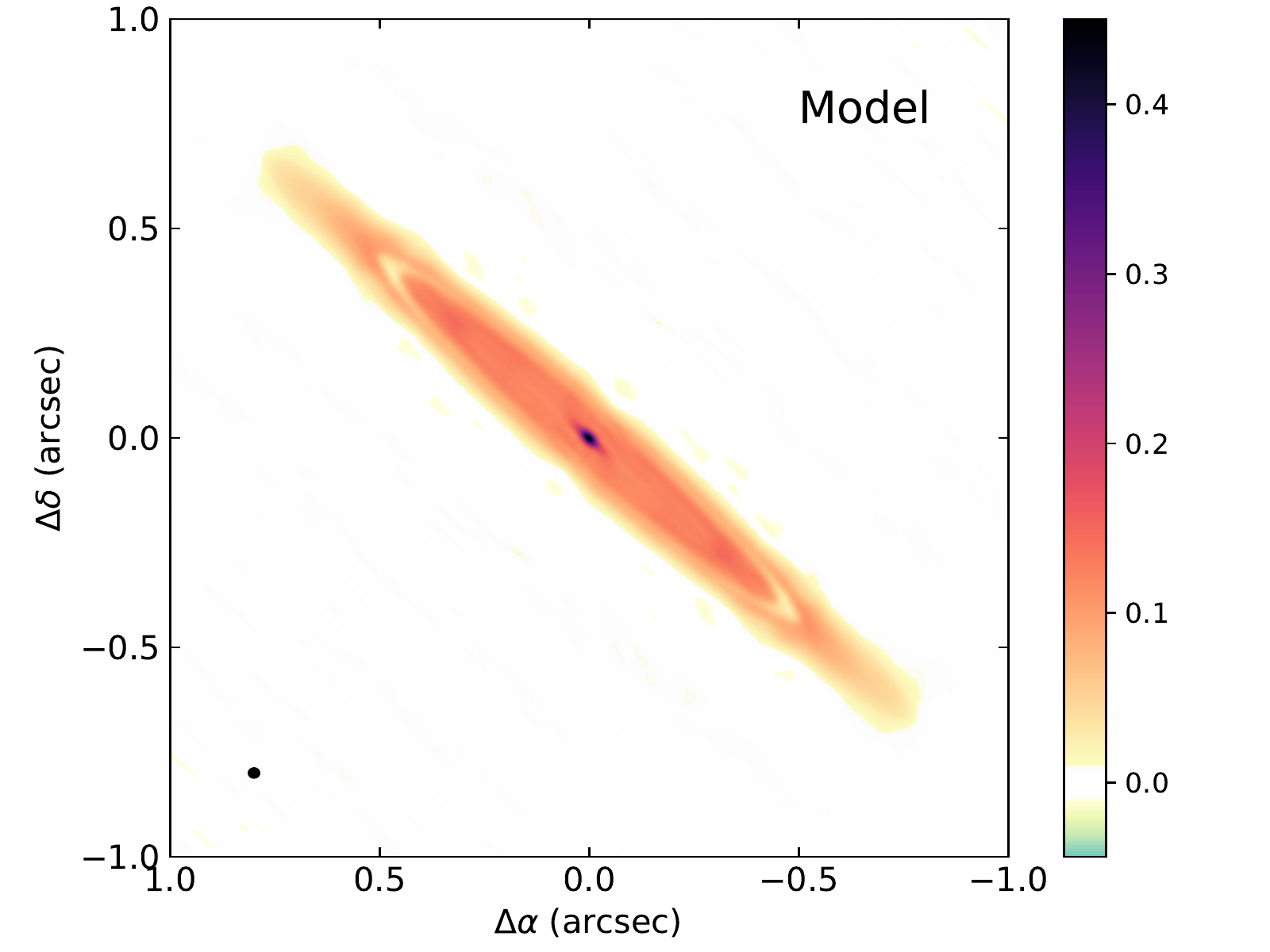}
    \includegraphics[width = 0.34\textwidth, trim={1cm 0cm 0.5cm 0cm}, clip]{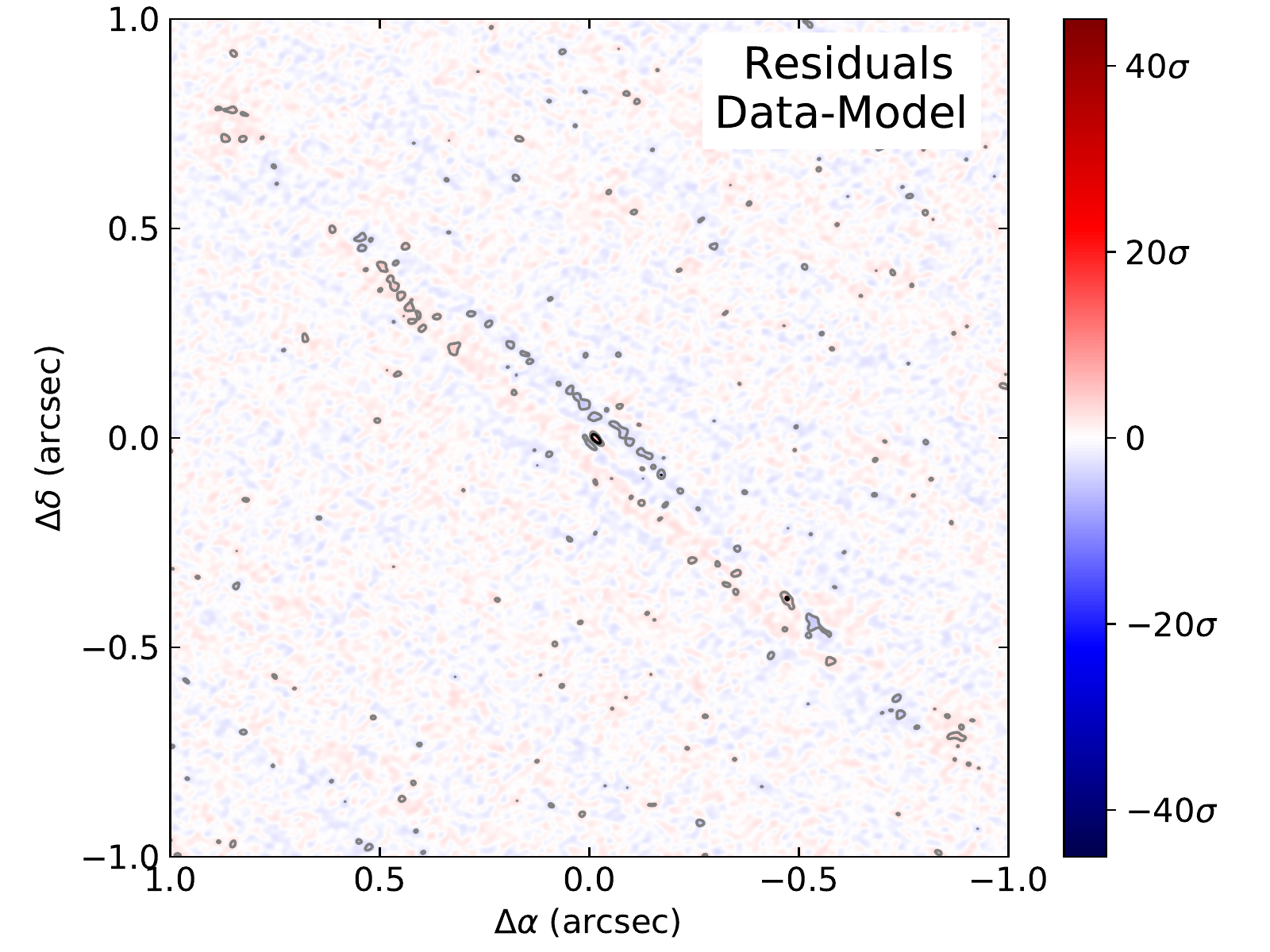}
    \caption{\emph{Left:} Continuum image of \Oph, \emph{Middle:} Synthetic observations of the model using the CASA simulator, \emph{Right:} Residual map. Deviations from 3$\sigma$ and 5$\sigma$ are indicated by black and grey contours.  In the scale bar $\sigma$ corresponds to the rms of the data image. The maximum of the scale corresponds to the peak signal-to-noise in the data ($\sim48.7\sigma$). }
    \label{fig:model_1330}
\end{figure*}

Previous modelings of \Oph\ by \citet{Wolff_2021} have demonstrated that some degree of vertical settling is needed to reproduce both the scattered light and millimeter images. In this  section and Sect.~\ref{sec:model_ALMA}, we mimic dust settling by considering two separated grain populations~\citep[e.g.,][]{Villenave_2019, Keppler_2018}. One layer is composed of large grains ($10-1000\ \mu$m) aimed to model the ALMA continuum map, and the other layer includes smaller dust particles ($0.01-10\ \mu$m) necessary to reproduce the SED and scattered light image.  
In addition, we note that we present a complementary model in Sect.~\ref{sec:settling}, using a different (parametric) settling prescription and considering a continuous distribution of dust particles. This second approach allows us to test the level of turbulence in the disk.

For the model with two layers of dust grains (this section and Sect.~\ref{sec:model_ALMA}), we build the layer of small grains based on the results of the comprehensive MCMC fitting presented in \citet{Wolff_2021}. We have used a model where  the small grains extend from 0.07~au to 170~au, they have a scale height of $H_{sd, 100au} = 9.7$~au, and a surface density and flaring exponents of -1 and 1.2, respectively. The parameters of the small grain layer are summarized in Table~\ref{tab:parameters}. They were chosen from the range of allowed parameters in \citet{Wolff_2021} and to provide a reasonable match of the SED and scattered light image (see Appendix~\ref{appdx:SED_HST}). We note that, for all regions of our model, we assumed that the gas is colocated with the dust with a gas-to-dust ratio of~100. 

For the new millimeter continuum data, our strategy followed an iterative process, looking for a representative model. We adjusted the parameters to reproduce both the surface brightness cut along the major axis and minimize the residual map, by way of visual inspection. 
In Sect.~\ref{sec:model_ALMA}, we focus on the radial structure and adopt a scale height for the large grains of $H_{ld, 100{\rm au}} = 0.5$\,au at 100\,au. Then, in Sect.~\ref{sec:vertical_extent}, we explore and discuss this scale height assumption in more details.

\subsection{ALMA continuum model}
\label{sec:model_ALMA}

In the right panel of Fig.~\ref{fig:cont_label}, we presented a labelled version of the continuum image of \Oph. We highlighted the main features that we aim to reproduce with radiative transfer modeling, namely ring~2, ring~1, the partially-depleted region~2, and the central region. To reproduce the ALMA continuum emission, we thus consider 5 regions in our model: (1) a central point source, representing the central peak; (2) a first ring, needed to reflect the elongated structure around the central point source, labeled region~1 in Fig~\ref{fig:cont_label}; (3) an inner region, to reproduce region~2; (4) a second ring, for ring~1, and (5) an outer ring, reproducing ring~2.

\begin{figure}
    \centering
    \includegraphics[width = 0.45\textwidth, trim={0.4cm 0cm 0.5cm 0cm}, clip]{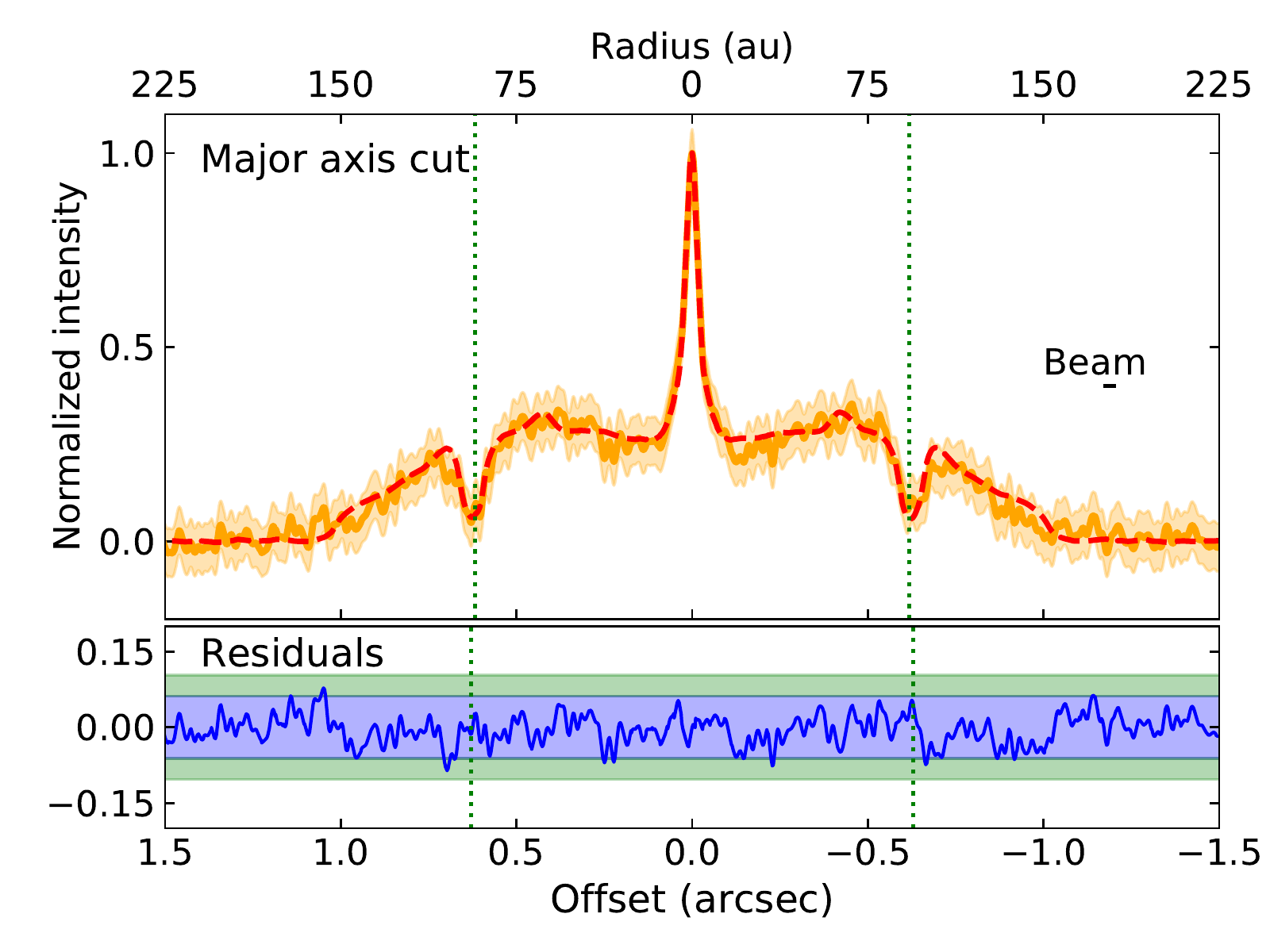}
    \caption{Cut along the major axis (top), with each map normalized to its maximum and the contours corresponding to 3$\sigma$ levels, and residual cut (bottom). The blue region in the residual map corresponds to 3$\sigma$ and the green to 5$\sigma$. Green vertical lines are at 0.63\arcsec.}
    \label{fig:model_1330_cut}
\medskip
    \includegraphics[width = 0.45\textwidth, trim={0.4cm 0cm 0.5cm 0cm}, clip]{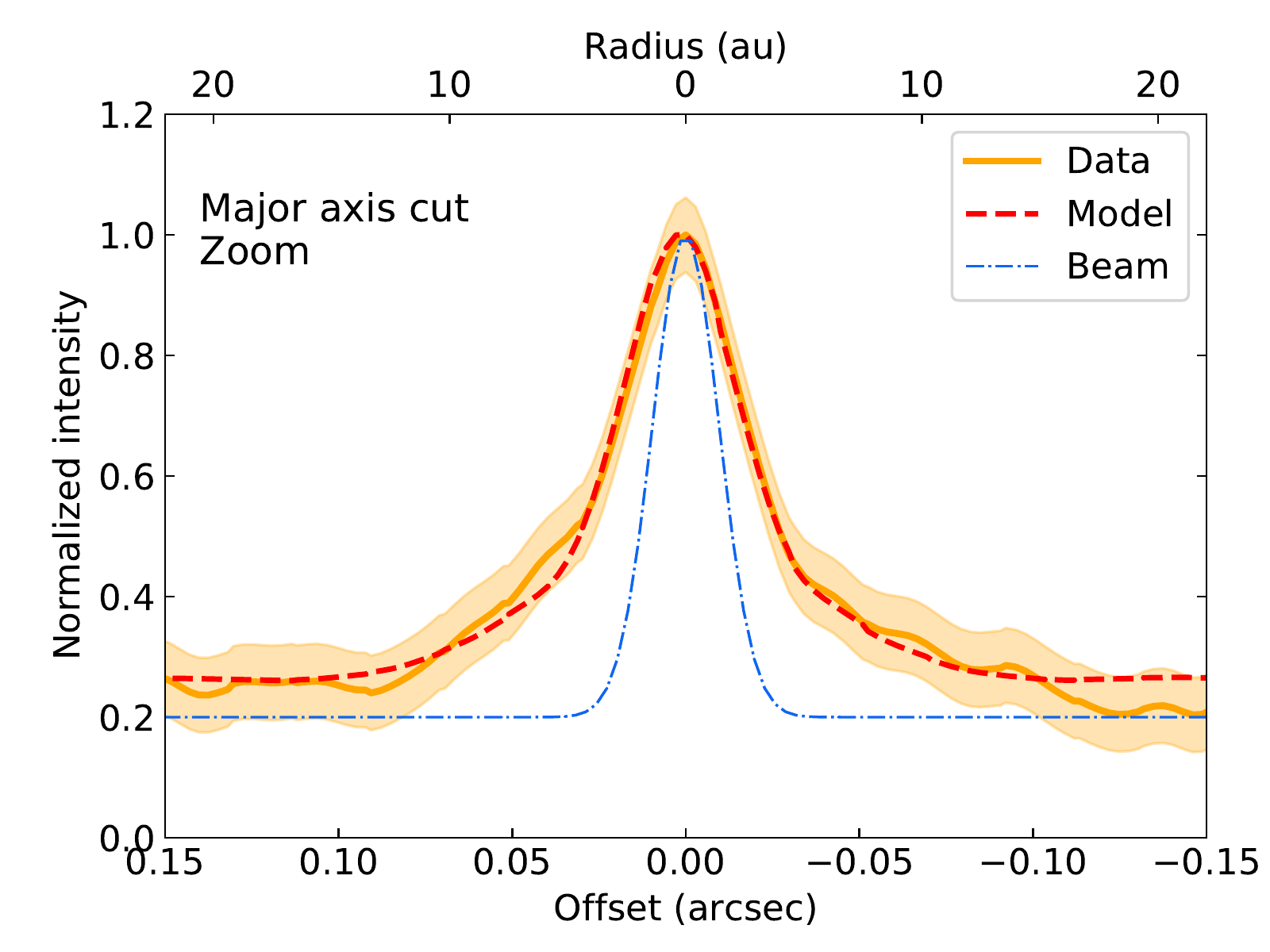}
    \caption{Zoom in the cut along the major axis (Fig.~\ref{fig:model_1330_cut}) to show the central region. We show the beam size by a dashed blue Gaussian.}
    \label{fig:model_1330_cut_zoom}
\end{figure}

In this section, we present the main characteristics of each region starting from the outside. Our strategy was to first fix the inner and outer radii of each region and then we followed an iterative process to estimate the best combination of dust masses in each region to reproduce the major axis cut and residual maps. We present the model in Fig.~\ref{fig:model_1330}, Fig.~\ref{fig:model_1330_cut}, and Fig.~\ref{fig:model_1330_cut_zoom}, and its parameters in Table~\ref{tab:parameters}.

\paragraph{Ring 2}
We reproduce the outer emission by implementing a broad ring between 98~au and 150~au. The brightness profile does not drop steeply at the edges of the disk, which we model with a steeper power-law exponent of $p= -6$.  Because this region is the best resolved, we used it to constrain the inclination of the system. We find that an inclination of $84\pm1^\circ$ matches best the major axis profile, the residual map, and the minor axis size of the disk. Both a higher and lower inclination lead to axis ratios which are not compatible with the data. We note that an inclination of 84$^\circ$ is consistent with the results of \citet{Wolff_2021}. We used this inclination and the observed position angle of the disk in the visibility fit presented in Sect~\ref{sec:frank}.

\begin{figure}
    \centering
    \includegraphics[width = 0.49\textwidth]{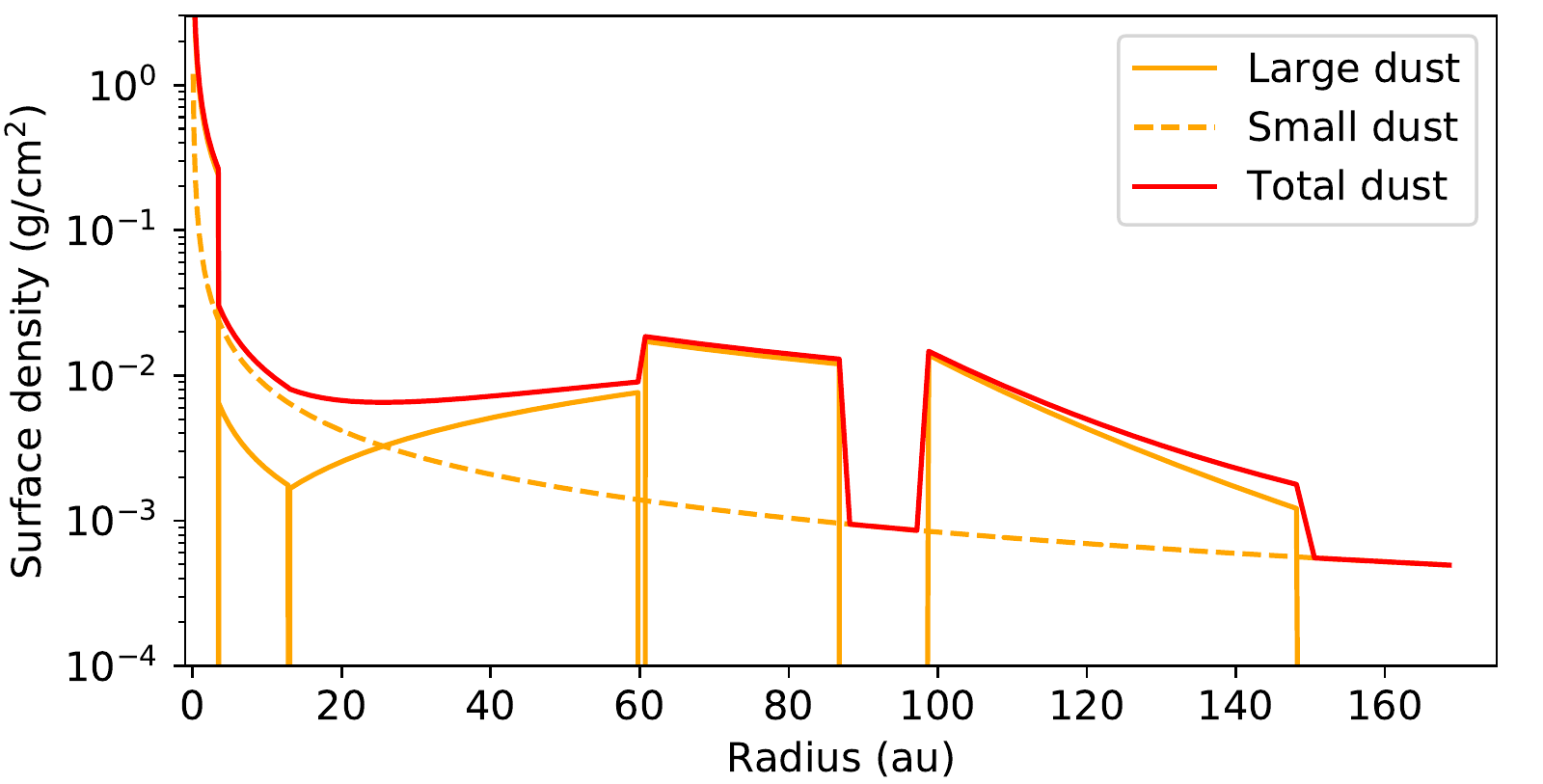}
    \caption{Dust surface densities of the model presented in Table~\ref{tab:parameters}.}
    \label{fig:surface_density}
\end{figure}

\paragraph{Ring 1}
The continuum millimeter image and \texttt{frank} profile (Fig.~\ref{fig:cont} and Fig.~\ref{fig:frank}) clearly show a ring centered at 73\,au ($\sim$ 0.5\arcsec). 
Because of the high inclination of the system, the ring is less clear in the major axis cut, as it is affected by projection effects. We reproduce this ring with a disk region between 60\,au and 87\,au.
After including ring~1 in our model, it became clear both by looking at the model images and the major axis cut that the region inner to it is not devoid of dust. To reproduce this feature, we introduced a region of large grains in region~2.

\begin{table}
\centering
\caption{Parameters for our radiative transfer model.}
\begin{tabular}{rlc}
\hline \hline
Inclination& ($^\circ$)& 84\\
PA& ($^\circ$)& 49 \\
\hline
&& \emph{ Central point source}\\
$a_\mathrm{min}-a_\mathrm{max}$&  ($\mu$m) &$10 - 1000$\\
$R_\mathrm{in}-R_\mathrm{out}$ & (au)&$0.07 - 3.5$\\
$M_\mathrm{dust}$&   (M$_\odot$)& $4\,\cdot\,$10$^{-6}$\\
H$_\mathrm{ld, 100au}$, $\beta$, p  &(au, $-$, $-$) & 0.5,  1.1, -1\\
\hline
&&\emph{Region 1}\\
$a_\mathrm{min}-a_\mathrm{max}$&  ($\mu$m) &$10 - 1000$\\
$R_\mathrm{in}-R_\mathrm{out}$ & (au)&$3.5 - 13$ \\
$M_\mathrm{dust}$&   (M$_\odot$)& $3\,\cdot\,$10$^{-7}$\\
H$_\mathrm{ld, 100au}$, $\beta$, p  &(au, $-$, $-$) & 0.5,  1.1, -1\\
\hline
&&\emph{Region 2}\\
$a_\mathrm{min}-a_\mathrm{max}$&  ($\mu$m) &$10 - 1000$\\
$R_\mathrm{in}-R_\mathrm{out}$ & (au)&$13 - 60$ \\
$M_\mathrm{dust}$&   (M$_\odot$)& $1.3\,\cdot\,$10$^{-5}$ \\
H$_\mathrm{ld, 100au}$, $\beta$, p  &(au, $-$, $-$) & 0.5,  1.1, +1\\
\hline
&& \emph{Ring 1}\\
$a_\mathrm{min}-a_\mathrm{max}$&  ($\mu$m) &$10 - 1000$\\
$R_\mathrm{in}-R_\mathrm{out}$ & (au)& $60 - 87$ \\
$M_\mathrm{dust}$&   (M$_\odot$)& $4\,\cdot\,$10$^{-5}$ \\
H$_\mathrm{ld, 100au}$, $\beta$, p  &(au, $-$, $-$) & 0.5,  1.1, -1\\
\hline
&& \emph{Ring 2} \\
$a_\mathrm{min}-a_\mathrm{max}$&  ($\mu$m) &$10 - 1000$\\
$R_\mathrm{in}-R_\mathrm{out}$ & (au)&$98 - $ 150\\
$M_\mathrm{dust}$&   (M$_\odot$)& $4\cdot10^{-5}$\\
H$_\mathrm{ld, 100au}$, $\beta$, p  &(au, $-$, $-$) & 0.5,  1.1, -6\\
\hline
&&\emph{Small grains}\\
$a_\mathrm{min}-a_\mathrm{max}$&  ($\mu$m)&$0.01 - 10$\\
$R_\mathrm{in}-R_\mathrm{out}$ & (au)&$0.07-170$ \\
$M_\mathrm{dust}$ & (M$_\odot$)& $2\,\cdot\,$10$^{-5}$\\
H$_\mathrm{sd, 100au}$, $\beta$, p  &(au, $-$, $-$) & 9.7,  1.2, -1\\
\hline
\end{tabular}
\tablecomments{Each parameter was adjusted during the modeling, except for the grain size ($a_\mathrm{min}-a_\mathrm{max}$) and the flaring exponent of each region ($\beta$). The parameters $H_{100au}$ and $\beta$ were defined in Eq.~(\ref{eq:scale_height}), and the surface density exponent in Eq.~(\ref{eq:surf_dens}).}
\label{tab:parameters}
\end{table}

\begin{figure*}
    \centering
    \includegraphics[width = \textwidth]{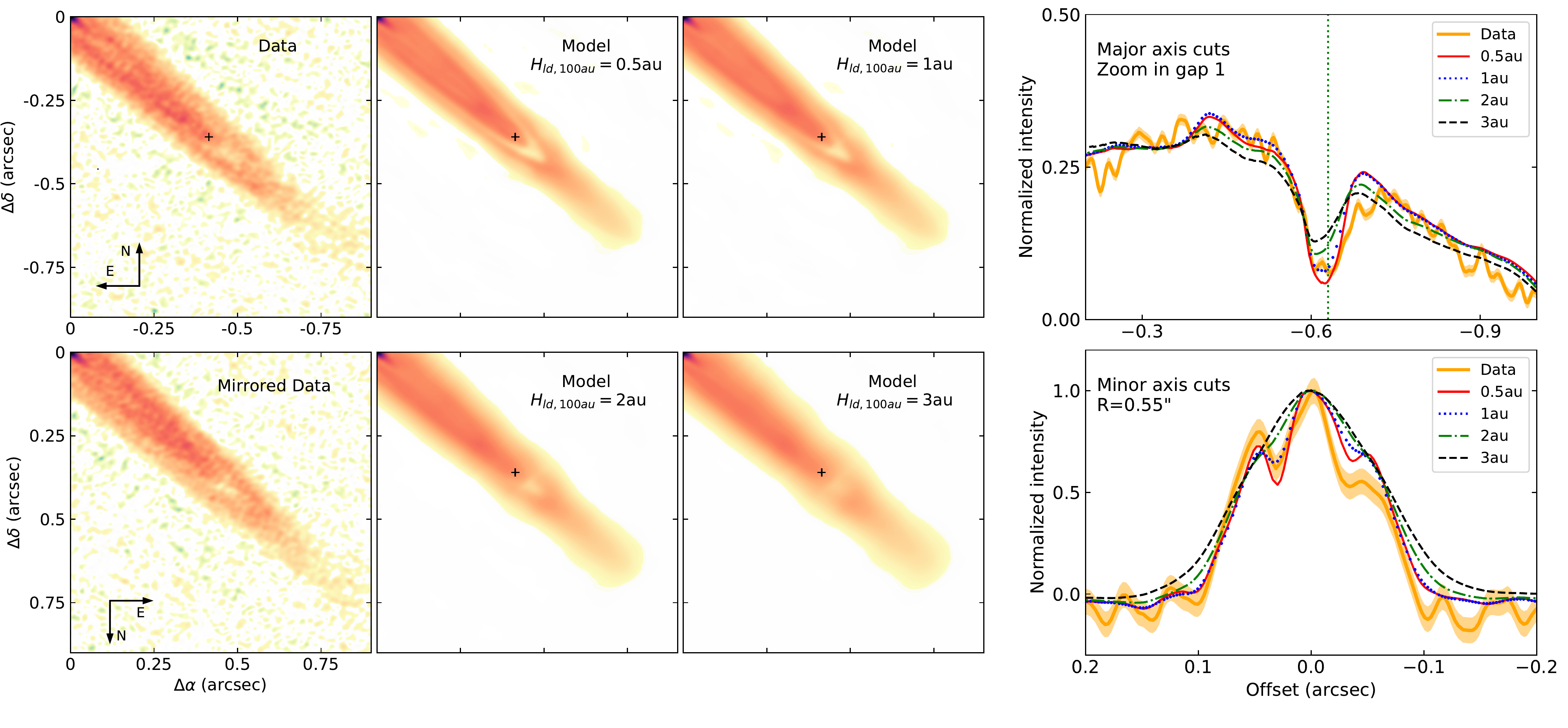}
    \caption{Scale height comparison. \emph{Top and bottom left panels:} Both sides of the continuum image of \Oph, \emph{Middle panels:} Model images of \Oph, with $H_{ld, 100{\rm au}}$ varying from 0.5 to 3 au. \emph{Top right panel:} Zoom in on gap 1~(0.63\arcsec\ is marked by the green vertical line) along the major axis of the disk, for the data (averaged from both sides) and models.  \emph{Bottom right panel:} Minor axis cuts of the data and the models, at 0.55\arcsec~(80 au) from the central point source, marked by a black cross on the data and model images. The uncertainty on the cuts corresponds to 1$\sigma$.}
    \label{fig:scale_height_comp}
\end{figure*}

\paragraph{Region~2}
To reproduce the increase of surface brightness with distance to the star in region~2 (between 0.09\arcsec and 0.45\arcsec), we included a disk region between 13\,au and 60\,au. However, based on the shape of the major axis profile, we found that a surface density exponent of $p=-1$ (as in most other disk regions) does not provide a good match to this region. With a decreasing surface density with radius, the profile of the region would not be flat but dropping steeply after a few au. Our results indicate that to reproduce the major axis profile between 13\,au and 60\,au (0.09\arcsec\ and 0.45\arcsec), region~2 needs to have an increasing surface density with radius, with $p=+1$. Such a profile is not expected in protoplanetary disks with a smooth pressure gradient and might indicate the presence of more complex structure, such as additional unresolved rings. We note that self-induced dust traps at the position of ring~1 might be able to reproduce a profile of increasing surface density with radius at millimeter wavelengths, corresponding to region~2~\citep{Vericel_2021, Gonzalez_2017}. 

\paragraph{Central point source and region 1}  
During our modeling, we found that the innermost region of \Oph\ can not be correctly modeled by an unique region extending from 0.07\,au to 13\,au. This is because, as seen in the zoomed major axis cut on Fig.~\ref{fig:model_1330_cut_zoom}, the central emission is not properly reproduced by an unresolved point source convolved with the corresponding beam size: it can be decomposed into two Gaussians with different widths. We reproduce the central region with a slightly resolved central point source, plus a second ring  extending up to 13\,au. 
We fixed the inner radius of the central point source to be a conservative estimate of the sublimation radius~\citep[see][their Section~3.2]{Wolff_2021}.\\

To summarize, we present the millimeter model images and model parameters in Fig.~\ref{fig:model_1330}, Fig.~\ref{fig:model_1330_cut}, Fig.~\ref{fig:model_1330_cut_zoom}, and Table~\ref{tab:parameters}. The SED and scattered light model are shown in Appendix~\ref{appdx:SED_HST}, and we also display the surface density of the model in Fig.~\ref{fig:surface_density} (see Appendix~\ref{appdx:surface_density} for details). The model has a total dust mass of  1.2$\,\cdot\,$10$^{-4}\ M_\odot$, which is in agreement with the results from MCMC modeling by \citet[][constraining $M_\mathrm{dust} >3 \cdot 10^{-5}\ M_\odot$ for the scattered light model]{Wolff_2021}. However, we note that with the current modeling the large grain layer is partially optically thick, which implies that the dust masses presented in Table~\ref{tab:parameters} are formally lower limits.

\section{Discussion}
\label{sec:discussion}

\subsection{Vertical extent of mm sized particles}
\label{sec:vertical_extent}

Despite the high inclination of \Oph, the new millimeter continuum image presented in this work reveals interesting radial substructures. Those can be used to constrain the vertical extent of large dust grains in this disk. In particular, the fact that the outer gap is well resolved readily indicates that the mm-emitting dust is constrained to a very thin midplane layer. 
In Sect.~\ref{sec:model_ALMA}, we presented a model with an extremely thin large grain scale height of $H_{ld, 100{\rm au}} =0.5$ au. Here, in order to constrain the vertical extent of millimeter dust particles, we follow the methodology of \citet{Pinte_2016} on HL~Tau and look for geometrical constraints on the vertical extent of the millimeter dust in the rings. We modify our model presented in Sect.~\ref{sec:model} to test larger scale heights for the large dust layers, with $H_{ld, 100{\rm au}} =$ 0.5, 1, 2, and 3 au at 100\,au. We keep the same disk radial structure, dust mass, and small grain layer as presented in Table~\ref{tab:parameters}, and only vary the value of $H_{ld, 100{\rm au}}$ in the large grains layers.   
In Fig.~\ref{fig:scale_height_comp}, we compare the data with the model images, a zoom in on the major axis cut at the position of the outer gap, and minor axis cuts across ring~1.

From the model images and major axis cut, one can clearly see that when $H_{ld, 100{\rm au}}$ increases, the outer gap gets filled in due to projection effects~\citep[see also][]{Pinte_2016}. 
By observing the major axis profiles displayed in the top right panel of Fig.~\ref{fig:scale_height_comp}, we find that the gap is clearly less deep for the model with $H_{ld, 100{\rm au}} = 3$\,au than in the data, which indicates that this model is too vertically thick to reproduce the observations.

In addition, we computed minor axis cuts at 0.55\arcsec ($\sim80$\,au) from the central point source along the major axis direction to assess how well the ring/gap is resolved in the data and in the model. In the observations, the cut along the minor axis at 0.55\arcsec\ reveals 3 peaks: one for each side of ring~2 and the central peak corresponding to ring~1. We also see a brightness asymmetry between the two peaks corresponding to the outer ring, which is mostly due to optical depth / geometrical effects. This asymmetry varies between 20\% and 40\% along the major axis.  
When compared to the models, we find that the models with a millimeter scale height of $H_{ld, 100{\rm au}} = 3$\,au and $H_{ld, 100{\rm au}} = 2$\,au do not show three peaks in the cut. Their minor axis profiles are smoother, with only one peak, which indicates that they are too thick vertically to reproduce the observations. On the other hand, the models with a scale height of $H_{ld, 100{\rm au}} = 1$ au and $H_{ld, 100{\rm au}} = 0.5$ au show the three peaks. The peaks are significantly clearer for the model with a large grain scale height of 0.5\,au, which provides the best match to the data. This indicates that (at least) the outer region of the disk is extremely thin vertically, with a vertical scale height of the order of $H_{ld, 100{\rm au}} \leq 0.5$\,au or less. 

Our results are in agreement with findings from \citet{Pinte_2016} who modeled the gaps and rings of HL\,Tau, and also with the conclusions of \citet{Villenave_2020} based on the comparison of the major and minor axis profiles of HK\,Tau\,B, HH\,30, and HH\,48 with some fiducial radiative transfer modeling. Both studies find a scale height of millimeter grains $H_{ld, 100{\rm au}}$ of about or less than one au in these systems. Our analysis of \Oph\ thus seems to generalize the finding that grains emitting at millimeter wavelengths are extremely thin vertically in the outer regions of protoplanetary disks to a larger number of disks, pointing to the possibility that this is the case in most protoplanetary disks. 
We discuss some implications of this result for the formation of wide-orbit planets in Sect.~\ref{sec:pa}.  

\vspace{1cm}

\subsection{Implications on the degree of dust-gas coupling}
\label{sec:alpha/St}

In the previous section, we have constrained the vertical extent of the millimeter dust particles to be $H_{ld, 100{\rm au}} \leq 0.5$\,au at 100\,au. This value is about one order of magnitude smaller than the small grains and gas scale height obtained by \citet{Wolff_2021} using MCMC fitting of the scattered light data (of $H_{g, 100{\rm au}} = H_{sd, 100{\rm au}} = 9.7 \pm3.5$\,au), and is indicative of efficient settling of millimeter grains in the disk. From the difference of scale height between the gas and millimeter dust grains, one can estimate the degree of coupling of this dust with gas, which is what we aim to do in this section. 
More specifically we characterize the ratio $\alpha/St$, where $\alpha$ represents the turbulence strength~\citep{Shakura_1973}, and $St$ is the Stokes number which describes the aerodynamic coupling between dust and gas.

In the classical 1D prescription of settling~\citep{Dubrulle_1995}, the vertical transport of particles is related to turbulence through a diffusion process. 
Assuming a balance between settling and turbulence, for grains in the Epstein regime ($St\ll1$), and under the assumption of $z\ll H_g$, the dust scale height can be written as follows~\citep[e.g.,][]{Youdin_2007, Dullemond_2018}:
\begin{equation}
    H_d = H_g \left(1 + \frac{StSc}{\alpha}\right)^{-1/2}
    \label{eq:dust_scale_height}
\end{equation}
where $Sc$ is the Schmidt number of the turbulence, characterizing the ratio of the turbulent viscosity over turbulent diffusivity. In this section, we follow \citet{Dullemond_2018} and \citet{Rosotti_2020} and assume that the Schmidt number of the turbulence is equal to $Sc=1$. 

Using Eq.~(\ref{eq:dust_scale_height}) for $H_{d} = H_{ld, 100au}=0.5$~au and $H_{g, 100au} = 6.2$~au (the lowest limit from \citealt{Wolff_2021}), we obtain an upper limit for the ratio $\alpha/St$ of $[\alpha/St_{ld}]_{100au}<6\cdot10^{-3}$, suggesting a strong decoupling between the large grains in this model and the gas in the vertical direction. 

This value is relatively small compared to previous estimates of the $\alpha/St$ ratio, which have been measured both in the vertical and in the radial directions. 
In this paragraph we compare our estimates with results from \citet{Doi_Kataoka_2021}, also studying the strength of the coupling in the vertical direction. \citet{Doi_Kataoka_2021} estimated $[\alpha/St]_{100au, \text{ HD}163296}<1\cdot10^{-2}$ for the outer ring of HD~163296 (and $[\alpha/St]_{67au, \text{ HD}163296}>2.4$ for the inner ring). Their relatively looser constraint on the coupling strength of the outer ring likely comes from the fact that HD~163296 is significantly less inclined than \Oph\ (47$^\circ$ vs 85$^\circ$) making the study of its vertical structure more difficult.

On the other hand,  \citet{Dullemond_2018} and \citet{Rosotti_2020} used the dust and gas width of several ringed systems to constrain the coupling strength in the radial direction.  They consistently obtained $[\alpha/St]_{100au, \text{ HD}163296}\geq4\cdot10^{-2}$ for the second ring of HD~163296, and $[\alpha/St]_{120au, \text{ AS}209}\geq0.13$ for AS~209, which is significantly higher than our estimate of the vertical coupling on \Oph\ and that of \citet{Doi_Kataoka_2021} in HD~163296.

In the radial direction, it is also clear that the rings in \Oph\ are not particularly thin (and do not appear Gaussian). In particular, there is only a factor $R_{CO, out}/R_{ld, out}\sim2.8$ in radial size between $^{12}$CO and the millimeter dust emission while the difference in the vertical direction reaches about ten times that value: $H_{g, 100au}/H_{ld, 100au}\sim 20$ (or at least $H_{g, 100au}/H_{ld, 100au}> 12$ if we assume the lowest limit from \citealt{Wolff_2021}). 
As previously suggested by \citet{Doi_Kataoka_2021}, these apparent inconsistencies might indicate that the turbulence level is different in the radial and vertical direction, or that the ring formation mechanism is different from what was assumed by \citet{Dullemond_2018} and \citet{Rosotti_2020}. The first possibility is also consistent with the results from \citet{Weber_2022}, who produced hydrodynamical simulations of V4046\,Sgr and found that,  to reproduce all their observations,  the turbulence in the vertical direction must be reduced compared to its value in the radial direction.

\subsection{Vertical settling}
\label{sec:settling}

The modeling presented in Sect.~\ref{sec:model_ALMA}, Sect.~\ref{sec:vertical_extent} and used in Sect.~\ref{sec:alpha/St} was based on the simplified assumption of two independent layers of dust particles, where the small grains ($<10\,\mu$m) have a large scale height and the large grains ($>10\,\mu$m) are concentrated into the midplane. We now aim to test a more realistic settling prescription (previously used by e.g., \citealt{Pinte_2016} and \citealt{Wolff_2021}), which allows to consider a continuous distribution of dust particles. 
We use the \citet{Fromang_Nelson_2009} prescription for settling (their equation 19), which implementation in \texttt{mcfost} is described in \citet{Pinte_2016}.  We assume that the gas vertical profile remains Gaussian and that the diffusion is constant vertically. With this prescription, each dust grain size follows its individual vertical density profile, which depends on the gas scale height, local surface density, and the turbulent viscosity coefficient $\alpha$~\citep{Shakura_1973}.  
This profile reproduces relatively well the vertical extent of dust grains obtained with global MHD simulations~\citep{Fromang_Nelson_2009}.  For completeness, we report the vertical density profile for a grain size $a$ that we adopt in this section:
\begin{equation}
    \rho(r, z, a)\propto \Sigma(r) \exp\left[-\frac{\Omega\tau_S(a)}{\tilde{D}}\left(e^{\frac{z^2}{2H_g^2}} -1\right) -\frac{z^2}{2H_g^2}\right]
    \label{eq:Fromang}
\end{equation}
with $\Omega=\sqrt{\frac{GM_\star}{r^3}}$, $\tilde{D} = \frac{H_g\alpha}{S_c\Omega}$, $\tau_S(a)=\frac{\rho_da}{\rho_gc_S}$, and where $c_S=H_g\Omega$ is the midplane sound speed, $H_g$ is the gas scale height, $S_c$ is the Schmidt number that is fixed to~1.5 in \texttt{mcfost}~\citep{Pinte_2016}, $\rho_d$ is the dust material density, and $\rho_g$ is the gas density in the midplane. We note that the Stokes number described in Sect.~\ref{sec:alpha/St} corresponds to $St = \Omega\tau_S(a)$. The degree of settling is set by varying the $\alpha$ parameter. We also note that when $z\ll H_g$, the dust density defined in Eq.~(\ref{eq:Fromang}) reduces to a simple Gaussian function with a scale height defined by Eq.~(\ref{eq:dust_scale_height})~\citep[e.g.,][]{Riols_2018}. 

The value of the turbulence parameter is usually predicted to range from $\alpha = 10^{-3} - 10^{-2}$\ when driven by the MRI (magnetorotational instability, e.g., \citealt{Balbus_1991}). On the other hand, in regions with low ionization and weak coupling between the gas and magnetic fields, recent studies have showed that hydrodynamic or non-ideal MHD effects may dominate, and could lead to turbulent parameter as low as $\alpha = 10^{-4} - 10^{-6}$~\citep{Flock_2020, Bai_2013}. 
In this context, we produced models for different settling strengths, obtained for an $\alpha$ parameter of $10^{-5}$, $10^{-4}$, and $10^{-3}$. 
For this modeling, we assume that the grain size integrated over the whole disk follows a simplified power law distribution as of $n(a)\mathrm{d}a \propto a^{-3.5}\,\mathrm{d}a$, with minimum and maximum grain sizes of 0.01 and 1000\,$\mu$m, respectively. We note that the power law size distribution may vary as a function of vertical height due to the presence of vertical settling in our model~\citep{Sierra_2020}. 
We distribute the grains over 5 radial regions coincident with the large grain regions presented in Table~\ref{tab:parameters}. All 5 regions are normalized to a gas scale height of $H_{g, 100{\rm au}} = 9.7$ au at 100~au with a flaring of $\beta =1.1$, but the large grains will have a smaller scale height than the gas because of the settling prescription considered ($H_{ld, 100{\rm au}}<H_{g, 100{\rm au}}$).   We adjust the mass of each region to obtain a model representing relatively well the millimeter image, and more specifically the levels of the surface brightness profile the major axis. The total dust mass in these models varies between 7.3 and 9.9\,$\cdot 10^{-5}\ M_\odot$.  As for Sect.~\ref{sec:model_ALMA}, we produced synthetic images of the models using the CASA simulator.

\begin{figure*}
    \centering
    \includegraphics[width =1 \textwidth]{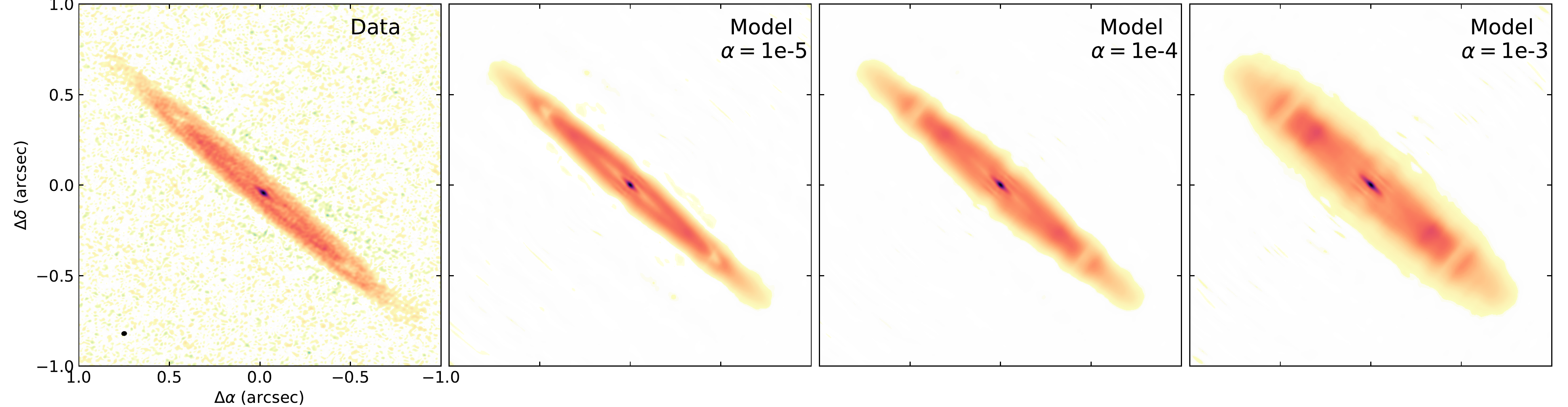}
    \caption{Data (left panel) and models using the settling prescription of \citet{Fromang_Nelson_2009}, with an $\alpha$ parameter of $10^{-5}$ (second left panel), $10^{-4}$ (second right panel), and $10^{-3}$ (right panel). }
    \label{fig:settling_fromang}
\end{figure*}

We present the 1.3\,mm model images obtained for an $\alpha$ parameter of $10^{-5}$, $10^{-4}$, and $10^{-3}$ in Fig.~\ref{fig:settling_fromang}.  For comparison with Sect.~\ref{sec:vertical_extent}, the scale height of 1~mm sized particles obtained in these models is $H_{1\text{mm}, 100au}= 0.95$~au, 2.7~au, and 5.7\,au, respectively.
Similarly to the previous section, the effect of vertical settling is clearly visible in these models. Focusing on the outer regions of the disk (ring~1, gap~1, and ring~2), we find that
the highest turbulence level leads to a high millimeter dust scale height, making the gap blurred due to projection effects. On  the other hand, the lowest turbulence tested of $\alpha =10^{-5}$ settles efficiently the large grains to the midplane and reproduces well the shape of the millimeter emission. 
When performing a similar study for various other gas scale heights, between $H_{g, 100{\rm au}} = 6.2$\,au and 13.2\,au, compatible with the range of posterior values estimated by~\citet[][]{Wolff_2021}, we also found that $\alpha =10^{-5}$ provides the best match with the observations.

Thus, we find that the millimeter observations of \Oph\ are consistent with a turbulent viscosity coefficient of $\alpha \leq10^{-5}$, at least in the outer regions of the disk (r $\sim$ 100 au).  
Together with results from turbulent line broadening in several protoplanetary disks~\citep[e.g.,][]{Flaherty_2020, Flaherty_2018, Flaherty_2017}, and of \citet{Pinte_2016} using a similar technique on HL\,Tau, our study suggests that the turbulence is low in the outer regions of protoplanetary disks. Additionnally, we note that when combining these results with the coupling estimate from Sect.~\ref{sec:alpha/St}, we find that the Stokes number of particles emitting at millimeter wavelengths must be greater than $St >1.7\cdot10^{-3}$ at 100\,au.

Finally, we note that the settling prescription implemented in this section depends on many parameters (e.g., grain size distribution, settling prescription, same turbulence in all the disk, constant grain opacity, ...) and that more complexity might need to be added in the models to be able to simultaneously reproduce the millimeter, scattered light images, and SED. Nevertheless, the high resolution datasets available for the highly-inclined disk in \Oph\ allows us to get independent measures of scale height for multiple dust sizes, which is extremely valuable to test and improve current dust settling models.

\subsection{Potential for wide-orbit planet formation in \Oph}
\label{sec:pa}

The exceptionally well-characterized outer disk of \Oph\ is strongly indicative of dust growth and settling. A significant fraction of the dust mass appears to have grown to near-mm sizes and to have settled towards a dense disk midplane layer, with $H_{ld, 100au}/H_{g, 100au}\sim~0.05$ at 100\,au. These two processes, dust growth and settling, are widely recognized as the key first steps for planetesimal formation and planetary growth \citep[for example, see reviews by][]{Youdin_2013,Johansen_2017,Ormel_2017}. Nevertheless, the outer parts of protoplanetary disks ($\gtrsim~50$\,au) are a challenging environment for planet formation due to inherently low solid surface densities and long orbital timescales. In this section, we address the potential for wide-orbit planet formation in \Oph, motivated by the ring-like features and dust depletion pattern that could be indicative of ongoing planet formation.
 
A common suggestion for the formation of wide-orbit giant planets is that such planets emerge through the direct fragmentation of young, massive, gravitationally unstable disks \citep{Helled_2014}. However, the apparent quiescent nature of \Oph\ with signs of extremely low of vertical stirring ($\alpha \lesssim 10^{-5}$) and little to no accretion onto the star~\citep{Flores_2021} rules out this scenario. Possibly, the formation of such gravitational-instability planets could have taken place earlier in the disk lifetime, but the early formation of such massive planets -- exceeding Jupiter in mass \citep{Kratter2010} -- would have resulted in a deep inner gas and dust cavities, that are not seen. Instead, as we argue below, planet formation could have proceeded more gently through core accretion via pebble accretion.
 
In the core accretion scenario \citep{Mizuno_1980,Pollack_1996}, the formation of a giant planet is initiated by the formation of a solid icy/rocky core, that accretes a H/He gaseous envelope of varying mass, resulting in either ice giants -- with an envelope mass smaller than the core mass -- or gas giants. However, the formation of a typical giant-planet core of about $10$ Earth masses (M$_\mathrm{E}$) cannot occur solely through the process of planetesimal accretion at orbits outside $10$\,au: formation times would exceed disk lifetimes even when assuming a planetesimal mass budget exceeding a solar dust-to-gas ratio by factor $10$ \citep{Rafikov_2004,Kobayashi_2010}. On the other hand, core growth can be accelerated by the direct accretion of pebbles, provided that, as is the case in \Oph, they are abundantly present and concentrated towards the disk midplane.
 
\begin{figure}
    \centering
    \includegraphics[width=0.45\textwidth]{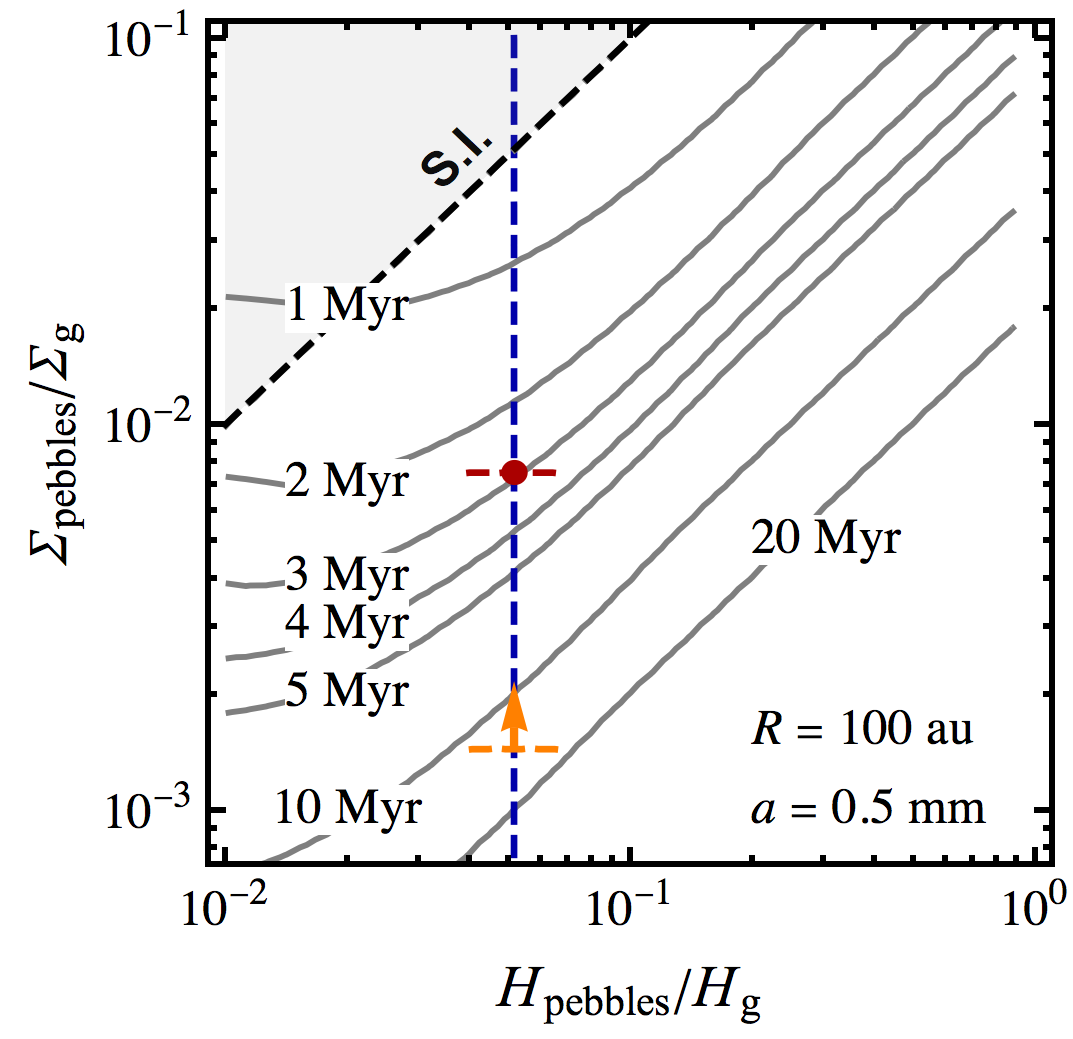}
    \caption{
    Time to grow a planetary embryo of 10$^{-2}$~M$_{\rm E}$ to a fully grown core of 10~M$_{\rm E}$ at $r=100$\,au, as given by the gray contours, depending on the pebble scale height and local pebble-to-gas surface density ratio, assuming pebbles $a=0.5$\,mm in size. The vertical blue dashed line gives the estimate for the pebble scale height in \Oph. The red point corresponds to a disk with solar dust-to-gas ratio, with a 75\%-mass fraction locked in pebbles. 
    The vertical orange dashed line corresponds to the assumption of massive gas disk, with a total mass of $0.1$\,M$_\odot$.
    The gray area shows where the midplane pebble-to-gas ratios is equal to, or above, unity, which corresponds to the parameter region in line with planetesimal formation through the streaming instability. %
    Pebble scale heights exceeding the one inferred here for \Oph\ would suppress planet formation in wide orbits.
    }
    \label{fig:pebble_accretion}
\end{figure}

The pebble accretion efficiency is highly dependent on the vertical pebble scale height, because only when the pebble accretion radius exceeds the pebble scale height, accretion reaches maximal efficiency  \citep[in the so-called 2D Hill accretion branch,][]{Lambrechts_2012}. Figure \ref{fig:pebble_accretion} illustrates the growth timescale as function of the pebble scale height and the pebble-to-gas ratio \citep[following][]{Lambrechts_2012,Lambrechts_2019}. 
Pebbles are assumed to be spherical with a radius of $a=0.5$\,mm.  We take for the pebble surface density the values inferred in Sect.~\ref{sec:model_ALMA} (see Table~\ref{tab:parameters} and Fig.~\ref{fig:surface_density}). However, we note that regions in the disk may be optically thick, which would result in higher dust-to-gas ratios than used here. We present two assumptions on the gas surface density. One where the dust-to-gas ratio is the nominal solar value of $0.01$, with a 75\%-mass fraction locked in pebbles (consistent with our model from Sect.~\ref{sec:model_ALMA}, see also Appendix~\ref{appdx:paparam}) and one where we assume a total disk mass of $0.1$\,M$_{\odot}$. 
The latter is an upper limit given that the gas disk is likely less massive: there is no strong gas accretion onto the star detected~\citep{Flores_2021} and no evidence for asymmetries in the disk~\citep[potentially triggered by gravitational instabilities, e.g.,][]{Hall_2019, Hall_2020}.
We choose a location of $r=100$\,au, corresponding to the inner edge of the outer ring 2.

The disk around \Oph\ appears to be conductive to core growth via pebble accretion within a range of reasonable values for the pebble-to-gas surface density, as shown in  Figure \ref{fig:pebble_accretion}. We find that a $10^{-2}$~M$_{\rm E}$ embryo can grow to up to 10~M$_{\rm E}$ at $60$\,au from the central star, within less than $10$\,Myr \citep[the maximum lifetime of protoplanetary disks,][]{Ribas_2015}.
Importantly, a pebble scale height of $H_{\rm pebbles}/H_g \sim 0.05$, 
consistent with $H_{ld, 100au}/H_{g, 100au}$ (see Sect. \ref{sec:model_ALMA} and \ref{sec:vertical_extent}, and also Appendix~\ref{appdx:surface_density}), 
strongly promotes core growth, while pebble scale heights that are a factor $10$ larger would largely suppress core growth by pebble accretion. 
Furthermore, in Appendix~\ref{appdx:paparam}, we show that core-growth timescales do not strongly depend on the chosen location. 
 We present results near the inner and outer edge of ring 1, at respectively $61$ and $85$\,AU. 
In contrast, our results do depend on the chosen particle size. In Appendix D, we illustrate that particles need to grow beyond $0.1$\,mm in size for pebble accretion to drive core growth, which is consistent with what is generally inferred in protoplanetary disks~\citep[e.g.,][]{Carrasco-Gonzalez_2019, Macias_2021, Tazzari_2021}.

A possible issue is that core growth by pebble accretion needs to be initiated by a sufficiently large seed mass that results from an episode of planetesimal formation.  Planetesimals are believed to be the results of the gravitational collapse of pebble swarms that self-concentrate through the streaming instability in the pebble midplane \citep{Youdin_2007,Johansen_2007}. The streaming instability requires dust growth to pebble sizes and a high degree of pebble settling (approaching midplane pebble-to-gas ratios near unity). 
Although planetesimal formation may be unlikely to be ongoing now  (see the gray upper triangle in Fig.~\ref{fig:pebble_accretion} marking where the midplane pebble-to-gas ratios is larger than one), the combined lack of strong vertical pebble stirring and the possibility of local moderate increases in pebble concentrations, would be conductive to planetesimal formation \citep{
Johansen_2009, Bai_2010, Carrera_2021}.
When planetesimal formation is triggered in simulations of the streaming instability, the analysis of the typical planetesimal mass distribution argues for large planetesimals in the outer disk \citep{Schafer_2017}. Following the mass scaling of \citet{Liu_2020}, we find planetesimals in the exponential tail of the mass distribution could reach masses exceeding 0.01~M$_{\rm E}$ beyond 60\,au (as assumed in Fig.\,\ref{fig:pebble_accretion}).
 
In summary, the low observed pebble scale height of the protoplanetary disk around \Oph\ is conducive to planetary growth by pebble accretion, even in wide orbits, exceeding $50$\,au. If indeed planet formation was initiated in these outer regions, it could possibly be consistent with the depletion of pebbles in the inner disk ($<60$\,au) and the ring-like features identified in this work. Thus, the \Oph-disk provides us with an exciting opportunity to study possibly ongoing planet formation at large orbital radii.

\section{Summary and conclusions}
\label{sec:concl}

In this paper, we present new high angular resolution ALMA continuum and $^{12}$CO millimeter images of the highly inclined disk \Oph. The $^{12}$CO image can be described by two distinct regions: 1) an inner region ($R\lesssim150$\,au) where the disk describes a clear X~shape, with a linear increase of the gas emission height with radius as of $z/r\sim0.3$, and 2) a flat outer region (for $R>200$\,au), with uniform brightness and temperature. In contrast with the $^{12}$CO emission, the scattered light surface brightness of  HST 0.6\,$\mu$m data appears flat at all radii. This indicates that scattered light comes from the outer regions of the disk, due to the high optical depth of micron-sized particles in \Oph. We find that the heights of scattered light and $^{12}$CO emission are similar at large radii, which suggests that both components are emitted at similar altitudes above the midplane.  
In addition, we find that the millimeter continuum emission from larger grains is less extended in both the radial and vertical direction compared to the scattered light and gas emission, which is in agreement with expectations from vertical settling and possibly radial drift.

Our millimeter continuum observations of \Oph\ reveal clear rings, which remained undetected with previous lower angular resolution observations. From the outside in, the disk shows two rings separated by a clear gap, some emission inside the first ring, and some bright central emission. 
We performed  comprehensive radiative transfer modeling of the ALMA continuum image in order to constrain the physical structure of the source. In particular, we used the resolved outer ring, located at $\sim$100\,au, to add strong geometrical constraints on the vertical extent of millimeter sized particles. Our modeling of \Oph\ indicates that the grains emitting at 1.3\,mm are extremely thin vertically, with a vertical scale height at 100\,au of $H_{ld, 100au}\leq 0.5$ au.
Because the vertical extent of small dust particles (and indirectly the gas) has been constrained to be $\sim10$~au at 100\,au~\citep{Wolff_2021}, this is a clear evidence that vertical settling is occurring in the disk.

Using a classical 1D prescription of settling and our estimate of dust and gas scale height, we estimate the degree of coupling of dust and gas to be $[\alpha/St]_{100au}<6\cdot10^{-3}$. This is value is particularly low compared to previous estimates in protoplanetary disks. 

We also aimed at constraining the turbulence parameter $\alpha$ in \Oph. To do so, we produced three additional radiative transfer models assuming the settling model from \citet{Fromang_Nelson_2009}, with $\alpha=10^{-3}, 10^{-4},\text{ and }10^{-5}$, and a gas vertical extent of 9.7\,au at 100\,au. 
We find that the coefficient of turbulent viscosity needs to be extremely low in the disk to reproduce the observations, of order of $\alpha \lesssim10^{-5}$ in the outer regions of the disk.

Finally, we used our results to test the pebble accretion scenario in the outer regions of the disk. The remarkably small pebble scale height of \Oph\ is particularly favorable for pebble accretion: we show a $10$ Earth mass planet can form in the outer disk, 
between approximately $60$ and $100$\,au, within less than 10~Myr.
If, on the other hand, the dust scale height would have been 10 times larger than our observational constraint, then this would largely suppress core growth by pebble accretion. 
Thus, the extreme vertical settling measured in \Oph\ can be the origin of the formation of wide-orbit planets. 
Further constraints on vertical settling in a larger number of protoplanetary disks might provide more insights into the dominant formation mechanism of wide-orbit planets.

\bigskip
\emph{Acknowledgments.} We thank the anonymous referee for their detailed revision of our work, which helped to improve the quality of this study. This paper makes use of the following ALMA data: ADS/JAO.ALMA\#2018.1.00958.S and 2016.1.00771.S. ALMA is a partnership of ESO (representing its member states), NSF (USA) and NINS (Japan), together with NRC (Canada), MOST and ASIAA (Taiwan), and KASI (Republic of Korea), in cooperation with the Republic of Chile. The Joint ALMA Observatory is operated by ESO, AUI/NRAO and NAOJ. The National Radio Astronomy Observatory is a facility of the National Science Foundation operated under cooperative agreement by Associated Universities, Inc. MV research was supported by an appointment to the NASA Postdoctoral Program at the NASA Jet Propulsion Laboratory, administered by Universities Space Research Association under contract with NASA. This project has received funding from the European Union's Horizon 2020 research and innovation programme under the Marie Sk\l{}odowska-Curie grant agreement Nº 210021. GD acknowledges support from NASA grants NNX15AC89G and NNX15AD95G/NExSS as well as 80NSSC18K0442. MB acknowledges funding from the European Research Council (ERC) under the European Union’s Horizon 2020 research and innovation programme (grant PROTOPLANETS No. 101002188). C.P. acknowledges funding from the Australian Research Council via FT170100040, and DP180104235. KRS acknowledges support from the NASA Exoplanet Exploration Program Office. AS acknowledges support from ANID/CONICYT Programa de Astronom\'ia Fondo ALMA-CONICYT 2018 31180052.

\bigskip

\emph{Software:} CASA \citep{McMullin_2007}, \texttt{mcfost} \citep{Pinte_2006, Pinte_2009}, \texttt{frank} \citep{Jennings_2020},  Matplotlib \citep{Hunter_2007}, Numpy \citep{Harris_2020},  \texttt{TinyTim} \citep{Krist_2011}.

\bigskip
 \copyright 2022. All rights reserved.





\appendix
\section{Tomographically Reconstructed Distribution}
\label{appdx:TRD}

\begin{figure*}
    \centering
    \includegraphics[width = 1\textwidth]{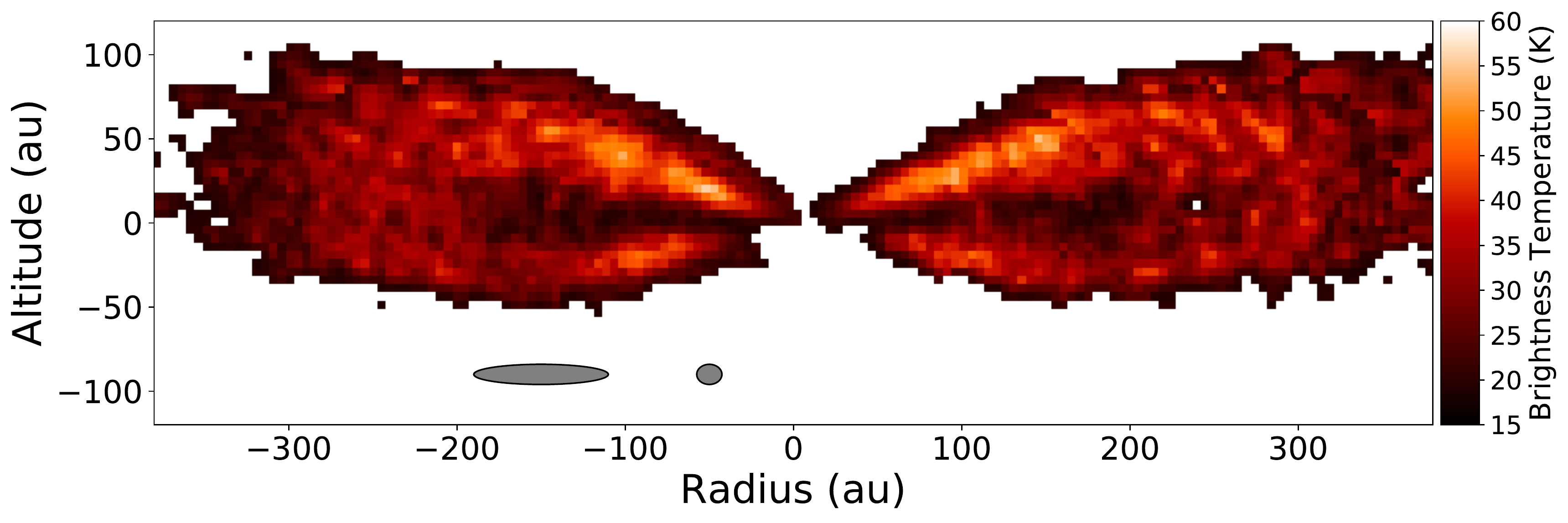}
    \caption{Tomographically reconstructed distribution of the $^{12}$CO. The reconstructed temperature maps are shown in physical units of radius and altitude, both in au. The grey ellipses shown on the bottom left part of the plot illustrate how the Keplerian velocity shear affects the resolution as a function of radius.} 
    \label{fig:TRD}
\end{figure*}

\citet{Flores_2021} first obtained a 2D temperature map of \Oph\ using the Tomographically Reconstructed Distribution method~\citep[TRD,][]{Dutrey_2017}. In Fig.~\ref{fig:TRD}, we present an updated temperature map of \Oph\ obtained  with the same technique from our higher angular resolution observations, after continuum subtraction. 
As in Sect.~\ref{sec:model}, we assumed a distance of 147~pc, a mass of 1.2~M$_\odot$, a beam size of 0.081\arcsec$\times$0.072\arcsec, position angle of 49$^\circ$, and an inclination angle of 84$^\circ$. Each pixel is about 5~au wide. 

Our new temperature map of \Oph\ presents the same global features than that previously obtained by \citet{Flores_2021} from lower angular resolution observations, but with significantly more details. In particular, the new map now clearly resolves the cold midplane in the inner regions ($R<200$\ au) of the disk. The new high-resolution data shows no evidence of a thin, cold mid-plane at radii beyond 200\,au. This is comparable with the outer radius of the millimeter dust, confirming that the shielded, high optical depth region of the disk ends at a radius of $\sim150-200$\,au.

The increased angular resolution also reduces the impact of beam smearing in the inner regions of the disk, which explains why peak temperatures almost twice as large are obtained with this new map compared to the previously published map. On the other hand, the reduced spectral resolution implies larger effect of Keplerian velocity shear at large radii compared to the previous observations (see radial extent of the ellipses in Fig.~\ref{fig:TRD}). As a reminder, we note that the Keplerian shear velocity is calculated as $dr=2rdv/v_K(r)$, where $v_K(r)$ is the Keplerian velocity of the disk, and $dv$ is the local line width. In our dataset, the local line width is dominated by the velocity resolution of 0.7\,km\,s$^{-1}$. 
In addition, we recover the isothermal region in the outer 200\ au of the disk. We refer to \citet{Flores_2021} for detailed analyze and interpretation of the full temperature structure of this disk. 

\section{SED and scattered light model image}
\label{appdx:SED_HST}

\begin{figure*}
    \centering
    \includegraphics[width = 0.29\textwidth, trim={2cm 0cm 2cm 0cm}, clip ]{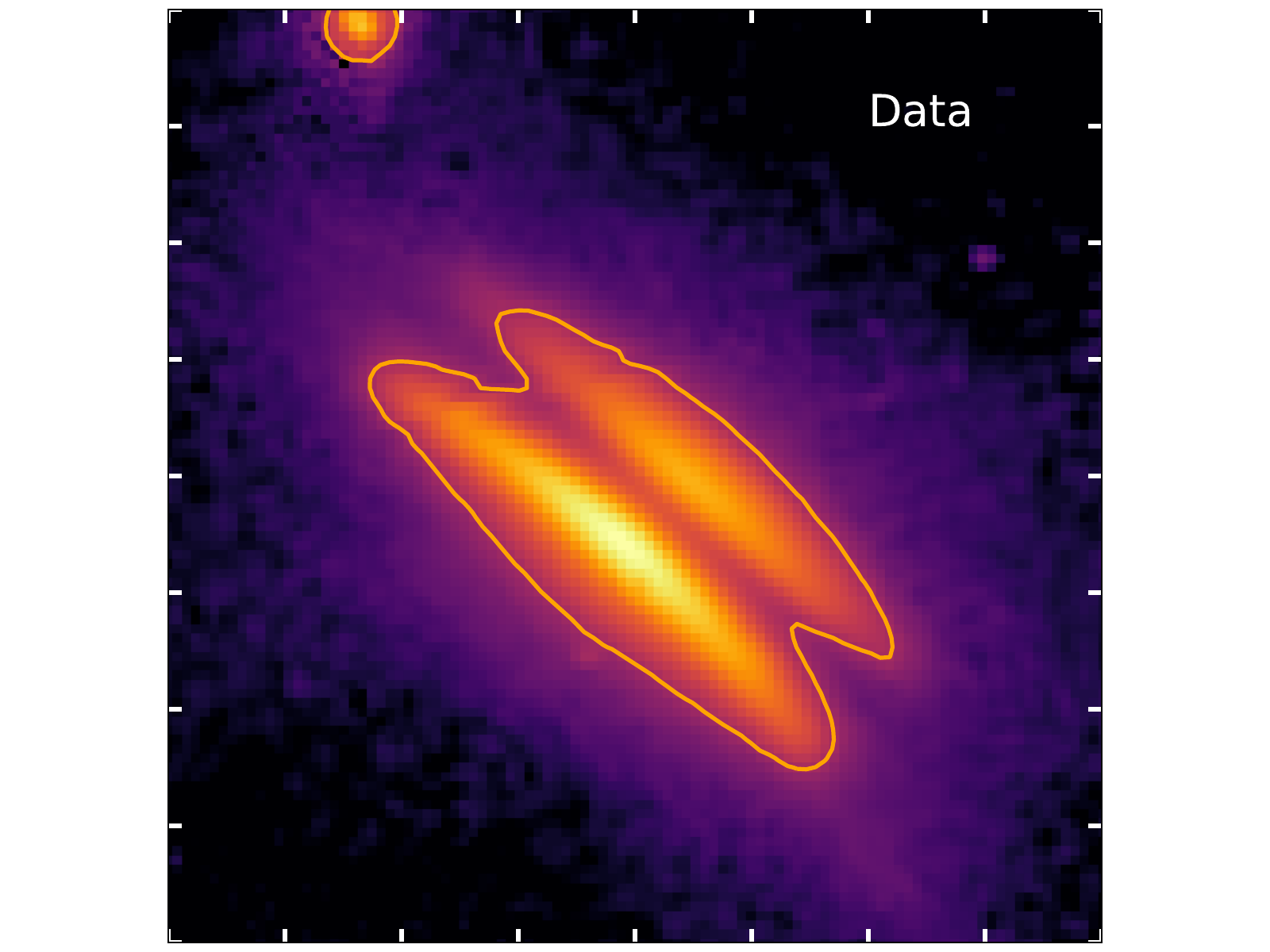}
    \includegraphics[width = 0.31\textwidth, trim={0cm 0cm 3.2cm 0cm}, clip ]{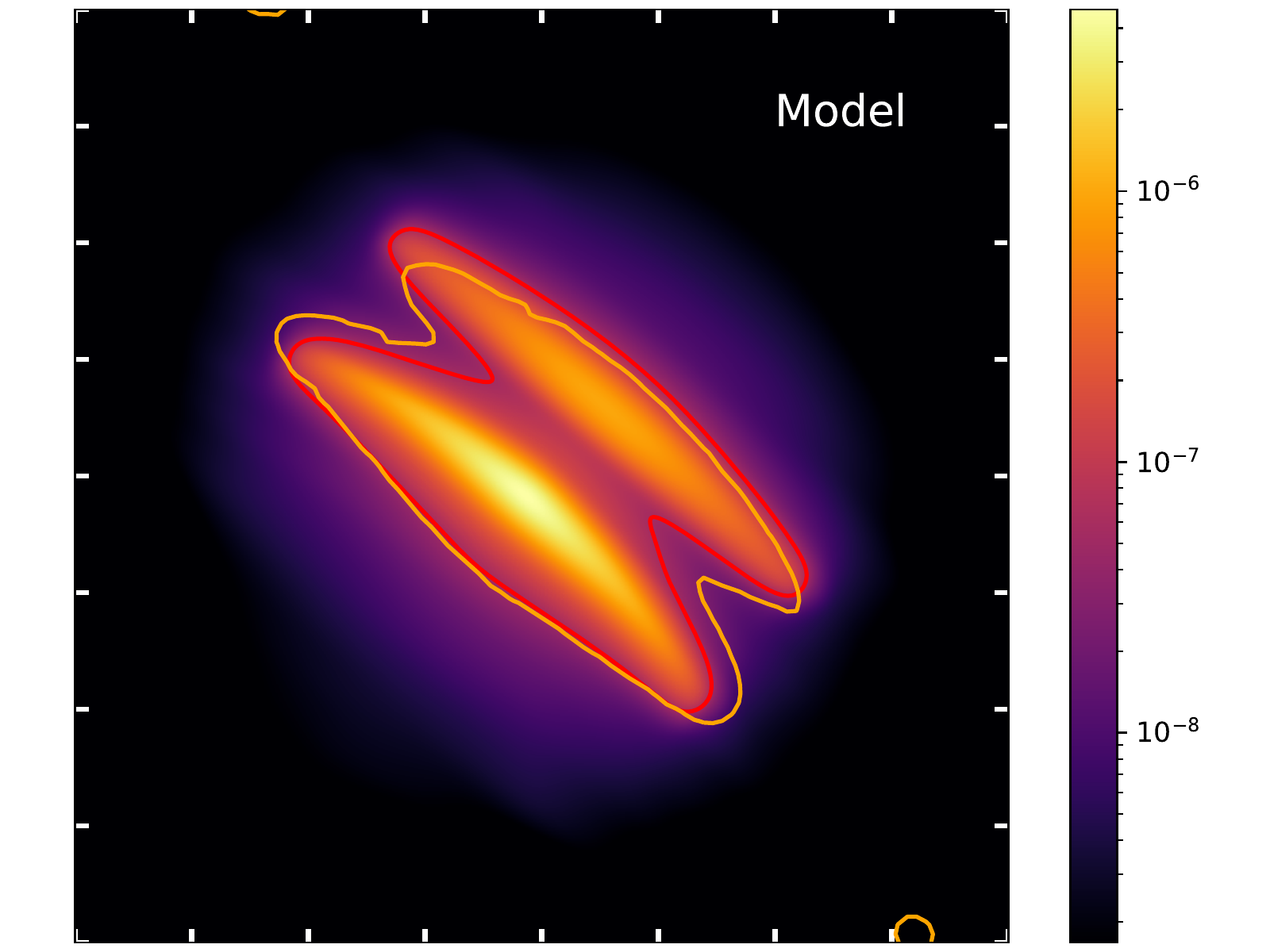}
    \includegraphics[width = 0.38\textwidth, trim={0.cm 0cm 0cm 0cm}, clip]{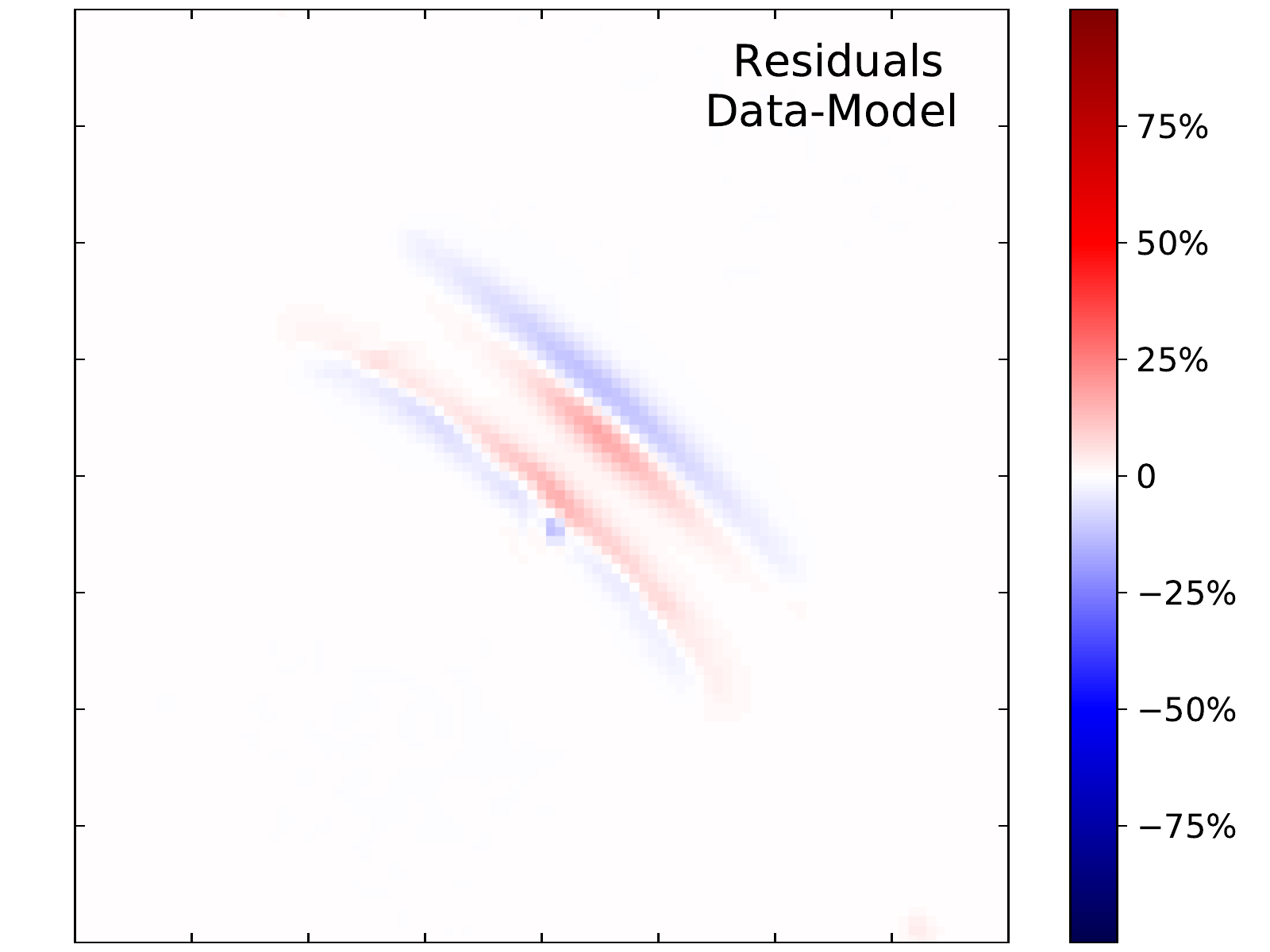}
    \caption{\emph{Left:} HST 0.6$\mu$m image, from \citet{Wolff_2021}, \emph{Middle:} 0.6 $\mu$m model obtained after convolution by a representative PSF. The x- and y-ticks indicate 0.5\arcsec. The yellow and red contours indicate 100$\sigma$ levels of the data and model, respectively.  \emph{Right:} Residual map.}
    \label{fig:model_0.6}
\end{figure*}

\begin{figure}
    \centering
    \includegraphics[width = 0.48\textwidth]{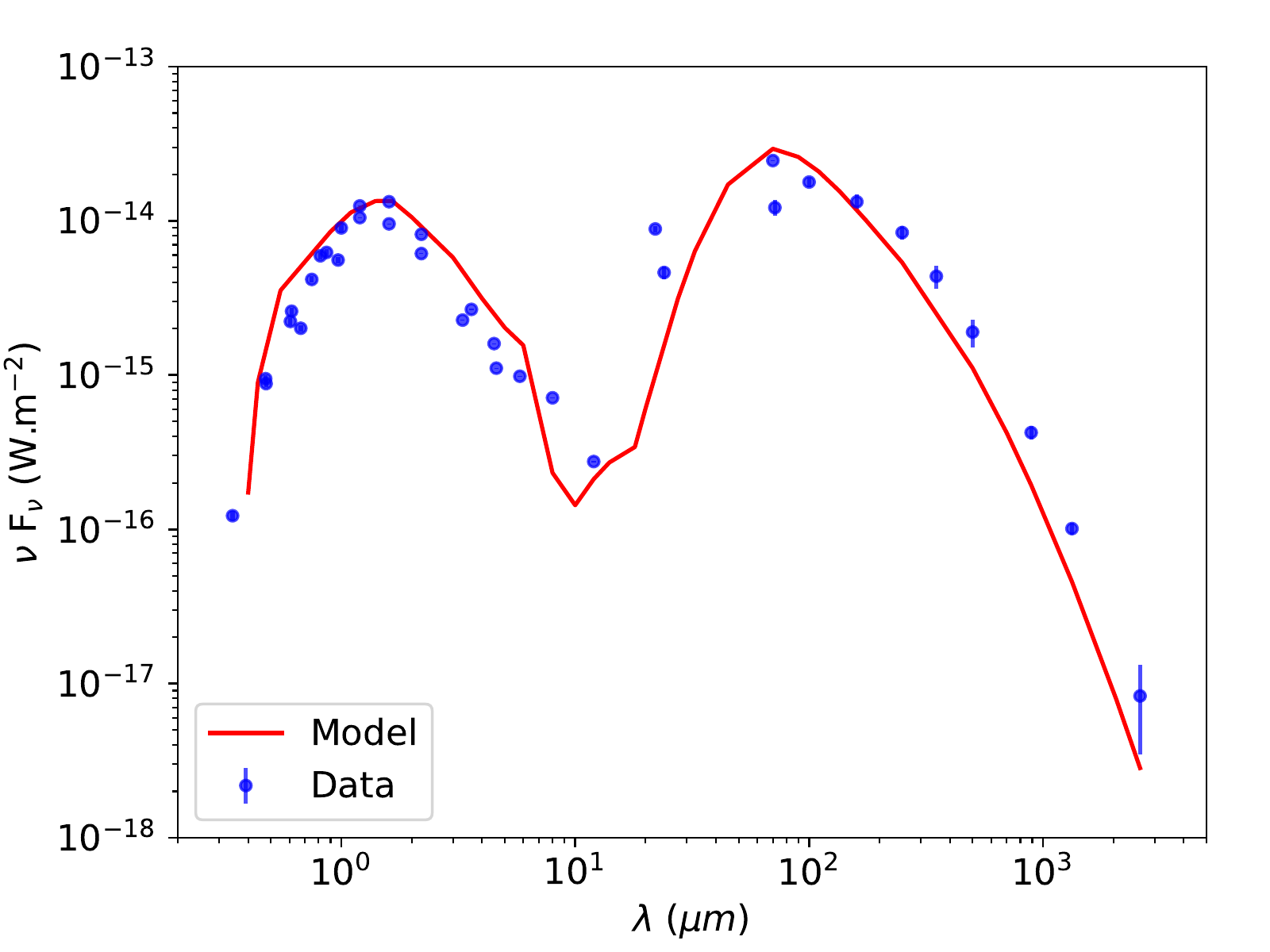}
    \caption{Spectral energy distribution of \Oph\,(blue circles) and our model prediction (red line). The model is corrected for interstellar extinction with an Av of 2 (see Sect.~\ref{sec:mcfost}).}
    \label{fig:SED}
\end{figure}

In Sect.~\ref{sec:model}, to  mimic dust vertical settling, we have added a layer of small dust particles located higher up in the disk than the large grain dust (see Table~\ref{tab:parameters}). This allows to obtain a relatively good agreement with the 0.6~$\mu$m maps and SED, which are presented in Fig.~\ref{fig:model_0.6} and Fig.~\ref{fig:SED}, respectively. We note that the photometry points used for the SED were obtained from \citet{Wolff_2021}. 

\section{Model surface density}
\label{appdx:surface_density}

In Fig.~\ref{fig:surface_density}, we presented the surface density of our model from Sect.~\ref{sec:model_ALMA}. We estimated it directly from our \texttt{mcfost} grid of the dust density (obtained with the command \texttt{-disk\_struct}) following:
\begin{equation}
    \text{Surface density(r)} = \sum_{z}\left( \sum_{a} \rho(a, r, z) \times m(a) \times z_{cell}(r) \right) 
\end{equation}
where $\rho(a, r, z)$ is the dust density of a particle of size $a$ at a radius $r$ and a height $z$, $m(a)$ is the mass of a particle of size $a$, and $z_{cell}(r)$ is the height of one \texttt{mcfost} cell at the radius $r$. 

We note that the gas surface density is not displayed in this plot, as we simply assumed a  gas-to-dust ratio of 100 in our modeling. In addition, the large dust is likely optically thick, which implies that the dust surface densities are formally lower limits.

\section{Pebble accretion timescales: parameter dependency}
\label{appdx:paparam}

\begin{figure}[h]
    \centering
    \includegraphics[height=0.25\textwidth]{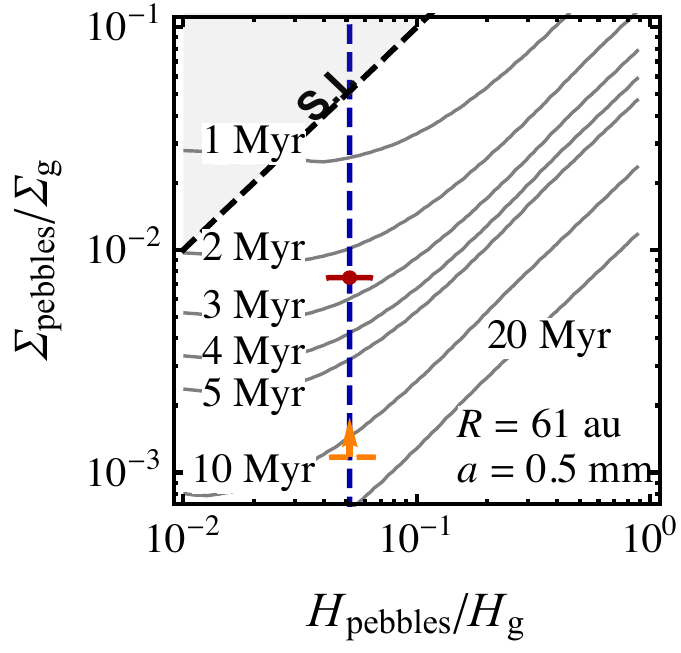}
    \includegraphics[height=0.25\textwidth]{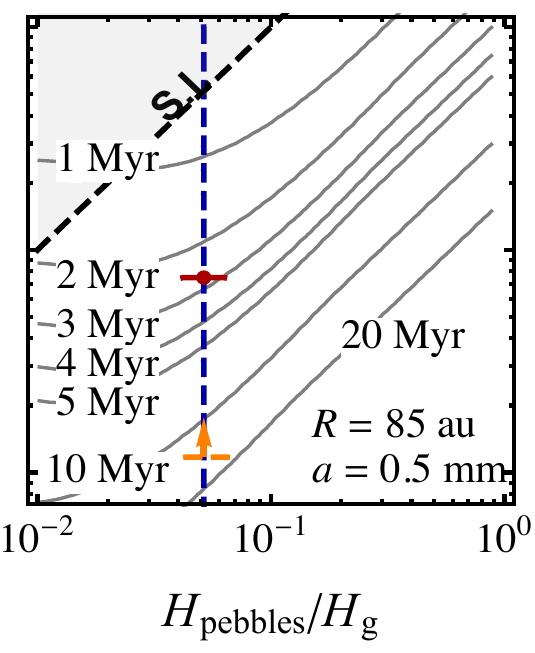}
    \caption{Dependency of the pebble accretion core-growth rates on the orbital radius. 
    The left panel shows results at $r=61$\,au and the right panel at $r=85$\,au, corresponding to the inner and outer edges of ring 1.}
    \label{fig:pa_semimajor}
\end{figure}

In this section, we explore the dependency of the core growth time scale on two model parameters: the orbital radius of the core and particle size. Results inside $100$\,au, for the inner and outer edge of ring~1 are shown in Fig.~\ref{fig:pa_semimajor}.
We have taken for the pebble surface densities the values shown in Fig.~\ref{fig:surface_density} at respectively $61$ and $85$\,au. 
 For the gas scale height we have used values based on the small dust component radius scaling presented in \citet{Wolff_2021}.
We furthermore have assumed that the pebble-to-gas scale height is constant with orbital radius, and set the value to the observed ratio of $H_{ld, 100au}/H_{g, 100au} = 0.05$. 
We do however caution that some disks, such as HD\,163296, have strong indications of radial changes in the pebble scale height between rings located in a similar orbital range between $60$ and $100$\,au \citep{Doi_Kataoka_2021}.

The weak orbital dependency in growth timescales, { from approximately $60$ to $100$\,au (see also Fig.\,\ref{fig:pebble_accretion}),} is the result of the here-assumed fixed particle size and surface density profile: the decrease in Stokes numbers on closer orbits reduces the accretion efficiency which is compensated by the increase in the pebble surface density.
We further note that in the $60$\,au region a planet with a mass exceeding $\sim 10$\,M$_{\rm E}$ can already accrete more than $50$\% of the mass flux of pebbles drifting inwards. This would cause a corresponding dip in the pebble surface density, similar to the one seen in Fig.\,\ref{fig:surface_density}, which could be further deepened as the planet reaches its pebble isolation mass~\citep{Lambrechts_2014,Bitsch_2018}.

\begin{figure}
    \centering
   \includegraphics[height=4.5cm]{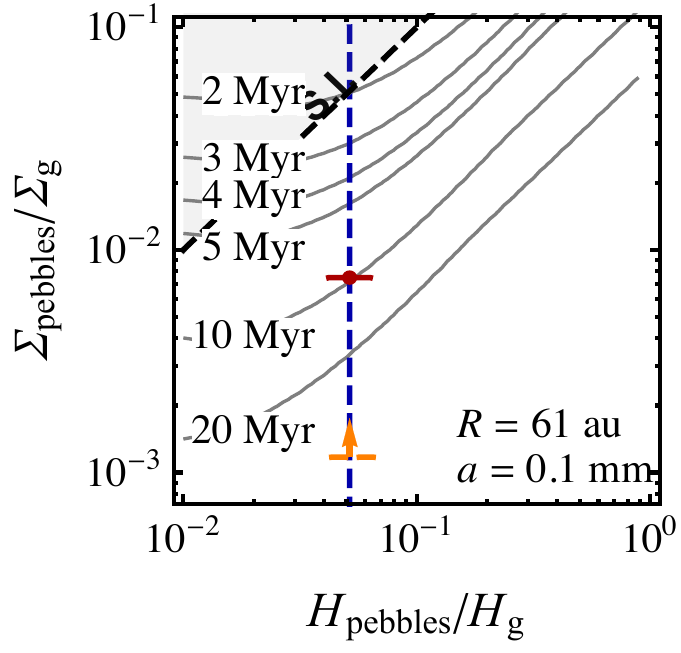}
   \includegraphics[height=4.5cm]{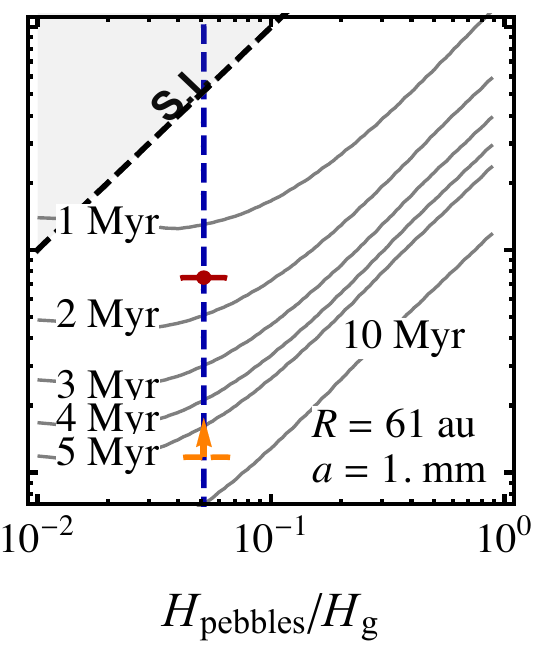}
    \caption{Dependency of the pebble accretion core-growth rates on the particle radius. The left panel shows results for pebbles $a=0.1$\,mm in size and the right panel for larger pebbles exclusively $a=1$\,mm in size.
    }
    \label{fig:pa_Rpebble}
\end{figure}

The model presented in Sect.~\ref{sec:model_ALMA} has a dust size distribution where $75$\,\% of the total mass in dust particles larger than 0.1\,mm. We therefore assume the pebble surface density to make up $75$\,\% of the total dust density. The exact mass fraction and upper limit of the  mass are however not fully constrained. We show in Fig.\,\ref{fig:pa_Rpebble}, our results when all pebbles have a size of $1$\,mm or $0.1$\,mm, 
{ covering the size range beyond particles can grow in observed protoplanetary discs~\citep[e.g.,][]{Tazzari_2021, Macias_2021, Carrasco-Gonzalez_2019}.
Core growth within disk lifetimes of approximately $10$\,Myr does require that dust particles grow beyond $0.1$\,mm in size, consistent with the here-presented model of the disk of \Oph.
}

In summary, we find that the timescale to grow an embryo to a $10$\,M$_{\rm E}$-core weakly depends on the orbital radius, but strongly on the maximal particle size for typical top-heavy particle size distributions.

\bibliography{sample631}{}
\bibliographystyle{aasjournal}



\end{document}